\documentclass[aps,prd,twocolumn,amsmath,amssymb,amsfonts,nofootinbib,superscriptaddress,altaffilletter]{revtex4}
\pdfoutput=1

\usepackage{graphicx}
\usepackage{hyperref}
\hypersetup{
  bookmarksopen=true,pdfauthor={LIGO and Virgo scientific collaborations},pdftitle={All-sky Search for Periodic Gravitational Waves in the Full S5 LIGO Data}
}
\usepackage{amssymb}
\usepackage{amsmath}
\usepackage{color}
\usepackage{supertabular}
\usepackage{dcolumn}

\def\EatH{Einstein@Home }
\def\sinc{\textrm{sinc}}
\def\ec{\textrm{~,}}
\def\ed{\textrm{~.}}
\def\fdot{f^{(1)}}

\newcommand{\SNR}{\textrm{SNR}}

\def\Re{\textrm{Re}}
\def\Im{\textrm{Im}}

\def\sci#1#2{#1\times10^{#2}}

\begin{document}

\title{ 
All-sky Search for Periodic Gravitational Waves in the Full S5 LIGO Data
}


\affiliation{LIGO - California Institute of Technology, Pasadena, CA  91125, USA$^\ast$}
\affiliation{California State University Fullerton, Fullerton CA 92831 USA$^\ast$}
\affiliation{SUPA, University of Glasgow, Glasgow, G12 8QQ, United Kingdom$^\ast$}
\affiliation{Laboratoire d'Annecy-le-Vieux de Physique des Particules (LAPP), Universit\'e de Savoie, CNRS/IN2P3, F-74941 Annecy-Le-Vieux, France$^\dagger$}
\affiliation{INFN, Sezione di Napoli $^a$; Universit\`a di Napoli 'Federico II'$^b$ Complesso Universitario di Monte S.Angelo, I-80126 Napoli; Universit\`a di Salerno, Fisciano, I-84084 Salerno$^c$, Italy$^\dagger$}
\affiliation{LIGO - Livingston Observatory, Livingston, LA  70754, USA$^\ast$}
\affiliation{Albert-Einstein-Institut, Max-Planck-Institut f\"ur Gravitationsphysik, D-30167 Hannover, Germany$^\dagger$}
\affiliation{Leibniz Universit\"at Hannover, D-30167 Hannover, Germany$^\ast$}
\affiliation{University of Wisconsin--Milwaukee, Milwaukee, WI  53201, USA$^\ast$}
\affiliation{Stanford University, Stanford, CA  94305, USA$^\ast$}
\affiliation{University of Florida, Gainesville, FL  32611, USA$^\ast$}
\affiliation{Louisiana State University, Baton Rouge, LA  70803, USA$^\ast$}
\affiliation{University of Birmingham, Birmingham, B15 2TT, United Kingdom$^\ast$}
\affiliation{INFN, Sezione di Roma$^a$; Universit\`a 'La Sapienza'$^b$, I-00185 Roma, Italy$^\dagger$}
\affiliation{LIGO - Hanford Observatory, Richland, WA  99352, USA$^\ast$}
\affiliation{Albert-Einstein-Institut, Max-Planck-Institut f\"ur Gravitationsphysik, D-14476 Golm, Germany$^\ast$}
\affiliation{Montana State University, Bozeman, MT 59717, USA$^\ast$}
\affiliation{European Gravitational Observatory (EGO), I-56021 Cascina (PI), Italy$^\dagger$}
\affiliation{Syracuse University, Syracuse, NY  13244, USA$^\ast$}
\affiliation{University of Western Australia, Crawley, WA 6009, Australia$^\ast$}
\affiliation{LIGO - Massachusetts Institute of Technology, Cambridge, MA 02139, USA$^\ast$}
\affiliation{Laboratoire AstroParticule et Cosmologie (APC) Universit\'e Paris Diderot, CNRS: IN2P3, CEA: DSM/IRFU, Observatoire de Paris, 10 rue A.Domon et L.Duquet, 75013 Paris - France$^\dagger$}
\affiliation{Columbia University, New York, NY  10027, USA$^\ast$}
\affiliation{INFN, Sezione di Pisa$^a$; Universit\`a di Pisa$^b$; I-56127 Pisa; Universit\`a di Siena, I-53100 Siena$^c$, Italy$^\dagger$}
\affiliation{Nikhef, Science Park, Amsterdam, the Netherlands$^a$; VU University Amsterdam, De Boelelaan 1081, 1081 HV Amsterdam, the Netherlands$^b$$^\dagger$}
\affiliation{The University of Texas at Brownsville and Texas Southmost College, Brownsville, TX  78520, USA$^\ast$}
\affiliation{San Jose State University, San Jose, CA 95192, USA$^\ast$}
\affiliation{Moscow State University, Moscow, 119992, Russia$^\ast$}
\affiliation{LAL, Universit\'e Paris-Sud, IN2P3/CNRS, F-91898 Orsay$^a$; ESPCI, CNRS,  F-75005 Paris$^b$, France$^\dagger$}
\affiliation{NASA/Goddard Space Flight Center, Greenbelt, MD  20771, USA$^\ast$}
\affiliation{The Pennsylvania State University, University Park, PA  16802, USA$^\ast$}
\affiliation{Universit\'e Nice-Sophia-Antipolis, CNRS, Observatoire de la C\^ote d'Azur, F-06304 Nice$^a$; Institut de Physique de Rennes, CNRS, Universit\'e de Rennes 1, 35042 Rennes$^b$, France$^\dagger$}
\affiliation{Laboratoire des Mat\'eriaux Avanc\'es (LMA), IN2P3/CNRS, F-69622 Villeurbanne, Lyon, France$^\dagger$}
\affiliation{Washington State University, Pullman, WA 99164, USA$^\ast$}
\affiliation{INFN, Sezione di Perugia$^a$; Universit\`a di Perugia$^b$, I-06123 Perugia,Italy$^\dagger$}
\affiliation{INFN, Sezione di Firenze, I-50019 Sesto Fiorentino$^a$; Universit\`a degli Studi di Urbino 'Carlo Bo', I-61029 Urbino$^b$, Italy$^\dagger$}
\affiliation{University of Oregon, Eugene, OR  97403, USA$^\ast$}
\affiliation{Laboratoire Kastler Brossel, ENS, CNRS, UPMC, Universit\'e Pierre et Marie Curie, 4 Place Jussieu, F-75005 Paris, France$^\dagger$}
\affiliation{Rutherford Appleton Laboratory, HSIC, Chilton, Didcot, Oxon OX11 0QX United Kingdom$^\ast$}
\affiliation{IM-PAN 00-956 Warsaw$^a$; Astronomical Observatory Warsaw University 00-478 Warsaw$^b$; CAMK-PAN 00-716 Warsaw$^c$; Bia{\l}ystok University 15-424 Bia{\l}ystok$^d$; IPJ 05-400 \'Swierk-Otwock$^e$; Institute of Astronomy 65-265 Zielona G\'ora$^f$,  Poland$^\dagger$}
\affiliation{University of Maryland, College Park, MD 20742 USA$^\ast$}
\affiliation{University of Massachusetts - Amherst, Amherst, MA 01003, USA$^\ast$}
\affiliation{The University of Mississippi, University, MS 38677, USA$^\ast$}
\affiliation{Canadian Institute for Theoretical Astrophysics, University of Toronto, Toronto, Ontario, M5S 3H8, Canada$^\ast$}
\affiliation{Tsinghua University, Beijing 100084 China$^\ast$}
\affiliation{University of Michigan, Ann Arbor, MI  48109, USA$^\ast$}
\affiliation{Charles Sturt University, Wagga Wagga, NSW 2678, Australia$^\ast$}
\affiliation{Caltech-CaRT, Pasadena, CA  91125, USA$^\ast$}
\affiliation{INFN, Sezione di Genova;  I-16146  Genova, Italy$^\dagger$}
\affiliation{Pusan National University, Busan 609-735, Korea$^\ast$}
\affiliation{Carleton College, Northfield, MN  55057, USA$^\ast$}
\affiliation{Australian National University, Canberra, ACT 0200, Australia$^\ast$}
\affiliation{The University of Melbourne, Parkville, VIC 3010, Australia$^\ast$}
\affiliation{Cardiff University, Cardiff, CF24 3AA, United Kingdom$^\ast$}
\affiliation{INFN, Sezione di Roma Tor Vergata$^a$; Universit\`a di Roma Tor Vergata, I-00133 Roma$^b$; Universit\`a dell'Aquila, I-67100 L'Aquila$^c$, Italy$^\dagger$}
\affiliation{University of Salerno, I-84084 Fisciano (Salerno), Italy and INFN (Sezione di Napoli), Italy$^\dagger$}
\affiliation{The University of Sheffield, Sheffield S10 2TN, United Kingdom$^\ast$}
\affiliation{RMKI, H-1121 Budapest, Konkoly Thege Mikl\'os \'ut 29-33, Hungary$^\dagger$}
\affiliation{INFN, Gruppo Collegato di Trento$^a$ and Universit\`a di Trento$^b$,  I-38050 Povo, Trento, Italy;   INFN, Sezione di Padova$^c$ and Universit\`a di Padova$^d$, I-35131 Padova, Italy$^\dagger$}
\affiliation{Inter-University Centre for Astronomy and Astrophysics, Pune - 411007, India$^\ast$}
\affiliation{University of Minnesota, Minneapolis, MN 55455, USA$^\ast$}
\affiliation{California Institute of Technology, Pasadena, CA  91125, USA$^\ast$}
\affiliation{Northwestern University, Evanston, IL  60208, USA$^\ast$}
\affiliation{The University of Texas at Austin, Austin, TX 78712, USA$^\ast$}
\affiliation{E\"otv\"os Lor\'and University, Budapest, 1117 Hungary$^\ast$}
\affiliation{University of Adelaide, Adelaide, SA 5005, Australia$^\ast$}
\affiliation{University of Szeged, 6720 Szeged, D\'om t\'er 9, Hungary$^\ast$}
\affiliation{Embry-Riddle Aeronautical University, Prescott, AZ   86301 USA$^\ast$}
\affiliation{National Institute for Mathematical Sciences, Daejeon 305-390, Korea$^\ast$}
\affiliation{Perimeter Institute for Theoretical Physics, Ontario, Canada, N2L 2Y5$^\ast$}
\affiliation{National Astronomical Observatory of Japan, Tokyo  181-8588, Japan$^\ast$}
\affiliation{Universitat de les Illes Balears, E-07122 Palma de Mallorca, Spain$^\ast$}
\affiliation{Korea Institute of Science and Technology Information, Daejeon 305-806, Korea$^\ast$}
\affiliation{University of Southampton, Southampton, SO17 1BJ, United Kingdom$^\ast$}
\affiliation{Institute of Applied Physics, Nizhny Novgorod, 603950, Russia$^\ast$}
\affiliation{Lund Observatory, Box 43, SE-221 00, Lund, Sweden$^\ast$}
\affiliation{Hanyang University, Seoul 133-791, Korea$^\ast$}
\affiliation{Seoul National University, Seoul 151-742, Korea$^\ast$}
\affiliation{University of Strathclyde, Glasgow, G1 1XQ, United Kingdom$^\ast$}
\affiliation{Southern University and A\&M College, Baton Rouge, LA  70813, USA$^\ast$}
\affiliation{University of Rochester, Rochester, NY  14627, USA$^\ast$}
\affiliation{Rochester Institute of Technology, Rochester, NY  14623, USA$^\ast$}
\affiliation{Hobart and William Smith Colleges, Geneva, NY  14456, USA$^\ast$}
\affiliation{University of Sannio at Benevento, I-82100 Benevento, Italy and INFN (Sezione di Napoli), Italy$^\ast$}
\affiliation{Louisiana Tech University, Ruston, LA  71272, USA$^\ast$}
\affiliation{McNeese State University, Lake Charles, LA 70609 USA$^\ast$}
\affiliation{University of Washington, Seattle, WA, 98195-4290, USA$^\ast$}
\affiliation{Andrews University, Berrien Springs, MI 49104 USA$^\ast$}
\affiliation{Trinity University, San Antonio, TX  78212, USA$^\ast$}
\affiliation{Southeastern Louisiana University, Hammond, LA  70402, USA$^\ast$}
\author{J.~Abadie$^\text{1}$}\noaffiliation\author{B.~P.~Abbott$^\text{1}$}\noaffiliation\author{R.~Abbott$^\text{1}$}\noaffiliation\author{T.~D.~Abbott$^\text{2}$}\noaffiliation\author{M.~Abernathy$^\text{3}$}\noaffiliation\author{T.~Accadia$^\text{4}$}\noaffiliation\author{F.~Acernese$^\text{5a,5c}$}\noaffiliation\author{C.~Adams$^\text{6}$}\noaffiliation\author{R.~Adhikari$^\text{1}$}\noaffiliation\author{C.~Affeldt$^\text{7,8}$}\noaffiliation
% ...
\collaboration{$^\ast$The LIGO Scientific Collaboration and $^\dagger$The Virgo 
Collaboration}\noaffiliation

\begin{abstract}
We report on an all-sky search for periodic gravitational waves in the frequency band 50-800~Hz and with the frequency time derivative in the range of $0$ through $\sci{-6}{-9}$~Hz/s. Such a signal could be produced by
a nearby spinning and slightly non-axisymmetric isolated neutron star in our galaxy. 
After recent improvements in the search program that yielded a 10$\times$ increase in computational efficiency,
we have searched in two years of data collected during LIGO's fifth science run and have obtained the most sensitive all-sky upper limits on gravitational wave strain to date. Near 150 Hz our upper limit on worst-case linearly polarized strain amplitude $h_0$ is  $\sci{1}{-24}$, while at the high end of our frequency range we achieve 
a worst-case upper limit of 
$\sci{3.8}{-24}$ 
for all polarizations and sky locations. 
These results constitute a factor of two improvement upon previously published data. A new detection pipeline utilizing a {\em Loosely Coherent} algorithm was able to follow up weaker outliers, increasing the volume of space where signals can be detected by a factor of 10, but has not revealed any gravitational wave signals. The pipeline has been tested for robustness with respect to deviations from the model of an isolated neutron star, such as caused by a low-mass or long-period binary companion.
\end{abstract}
%
%
\maketitle

\section{Introduction}
\label{sec:introduction}

In this paper we report the results of an all-sky search for continuous, nearly monochromatic gravitational waves on data from LIGO's fifth science (S5) run. The search covered frequencies from 50~Hz through 800~Hz and frequency derivatives from 0 through $\sci{-6}{-9}$~Hz/s.

A number of searches have been carried out previously in LIGO data~\cite{S4IncoherentPaper, EarlyS5Paper, S2TDPaper, S3S4TDPaper, S2FstatPaper, Crab, pulsars3, CasA}, including coherent searches for gravitational radiation from known radio and X-ray pulsars. An \EatH search running on the BOINC infrastructure \cite{BOINC} has performed blind all-sky searches on S4 and S5 data \cite{S4EH, S5EH}.

The results in this paper were produced with the PowerFlux search code. It was first described in \cite{S4IncoherentPaper} together with two other semi-coherent search pipelines (Hough, Stackslide). The sensitivities of all three methods were compared, with PowerFlux showing better results in
frequency bands lacking severe spectral artifacts.
A subsequent article~\cite{EarlyS5Paper} based on the first eight months of data from the S5 run featured improved upper limits and an opportunistic detection search. 

The analysis of the full data set from the fifth science run described in this paper has several distinguishing features from previously published results:
\begin{itemize}
 \item The data spanning two years of observation is the most sensitive to date. In particular, the intrinsic detector sensitivity in the low-frequency region of 100-300~Hz (taking into account integration time) will likely not be surpassed until advanced versions of the LIGO and Virgo interferometers come into operation. 

 \item The large data volume from the full S5 run required a rework of the PowerFlux code, resulting in a factor of 10 improvement in speed when iterating over multiple values of possible signal frequency derivative, while reporting more detailed search results. That partially compensated for the large factor in computational cost
   incurred by analyzing a longer time span, allowing frequencies up to
   800 Hz to be searched in a reasonable amount of time.
   The range of (negative) frequency derivatives considered, as large in
   magnitude as $\sci{-6}{-9}$~Hz/s, was slightly wider than in the
   previous search \cite{EarlyS5Paper}.
   Thus, this new
   search supersedes the previous search results up to 800 Hz.

  \item The detection search has been improved to process outliers down to signal-to-noise ratio $\SNR \ge 7$ using data from both the H1 and L1 interferometers. The previous search \cite{EarlyS5Paper} rejected candidates with combined $\SNR \le 8.5$. The new lower threshold is at the level of Gaussian noise, and new techniques were used to eliminate random coincidences.

  \item The followup of outliers employs the new {\em Loosely Coherent} algorithm \cite{loosely_coherent}.

\end{itemize}

We have observed no evidence of gravitational radiation and have established the most sensitive upper limits to date in the frequency band 50-800 Hz. Near 150 Hz our strain sensitivity to 
a neutron star with the most unfavorable sky location and orientation (``worst case'') yields a 95\% confidence level
upper limit of $\sci{1}{-24}$, while at the high end of our frequency range we achieve 
a worst-case upper limit of 
$\sci{3.8}{-24}$.

\section{LIGO interferometers and S5 science run}

The LIGO gravitational wave network consists of two observatories, one in Hanford, Washington and the other in Livingston, Louisiana, separated by a 3000~km baseline. During the S5 run each site housed one suspended interferometer with 4~km long arms. In addition, the Washington observatory housed a less sensitive 2~km interferometer, the data from which was not used in this search.

The fifth science run spanned a nearly two-year period of data acquisition. This analysis used data from GPS 816070843 (2005 Nov 15 06:20:30 UTC) through GPS 878044141 (2007 Nov 02 13:08:47 UTC). Since interferometers sporadically fall out of operation (``lose lock'') due to environmental or instrumental disturbances or for scheduled maintenance periods, the dataset is not contiguous. The Hanford interferometer H1 had a duty factor of 78\%, while the Livingston interferometer L1 had a duty factor of 66\%. The sensitivity was not uniform, exhibiting a $\sim 10$\% daily variation from anthropogenic activity as well as gradual improvement toward the end of the run \cite{LIGO_detector, S5_calibration}.

\section{The search for continuous gravitational radiation}
The search results described in this paper assume a classical model of a spinning neutron star with a fixed, asymmetric second moment that produces circularly polarized gravitational radiation along the rotation axis and linearly polarized radiation in the directions perpendicular to the rotation axis.
The assumed signal model is thus
\begin{equation}
\begin{array}{l}
h(t)=h_0\left(F_+(t, \alpha, \delta, \psi)\frac{1+\cos^2(\iota)}{2}\cos(\Phi(t))+\right.\\
\quad\quad\quad \left.\vphantom{\frac{1+\cos^2(\iota)}{2}}+F_\times(t, \alpha, \delta, \psi)\cos(\iota)\sin(\Phi(t))\right)\ec
\end{array}
\end{equation}

\noindent where $F_+$ and $F_\times$ characterize the detector responses to signals with ``$+$'' and ``$\times$'' 
quadrupolar polarizations, the sky location is described by right ascension $\alpha$ and declination $\delta$, $\iota$ describes the inclination of the source rotation axis to the line of sight, and the phase evolution of the signal is given by the formula
\begin{equation}
\label{eqn:phase_evolution}
\Phi(t)=2\pi(f_\textrm{source}(t-t_0)+\fdot(t-t_0)^2/2)+\phi\ec
\end{equation}
with $f_\textrm{source}$ being the source frequency and $\fdot$ denoting the first frequency derivative (for which we also use the shorter term {\em spindown}). $\phi$ denotes the initial phase with respect to reference time $t_0$. $t$ is time in the solar system barycenter frame. When expressed as a function of local time of ground-based detectors it includes the sky-position-dependent Doppler shift.
We use $\psi$ to denote the polarization angle of projected source rotation axis in the sky plane.

Our search algorithms calculate power for a bank of such templates and compute upper limits and signal-to-noise ratios for each template based on comparison to templates with nearby frequencies and the same sky location and spindown.

The search proceeded in two stages. First, the main {\em PowerFlux} code was run to establish upper limits and produce lists of outliers with signal-to-noise ratio (SNR) greater than 5. Next, the {\em Loosely Coherent} pipeline was used to reject or confirm collected outliers.

The upper limits are reported in terms of the worst-case value of $h_0$ (which applies to linear polarizations with $\iota=\pi/2$) and for the most sensitive circular polarization ($\iota=0$~or~$\pi$).
The pipeline does retain some sensitivity, however, to more general GW polarization models, including 
a longitudinal component, and to slow amplitude evolution.

\begin{figure*}[htbp]
\begin{center}
  \includegraphics[width=7.2in]{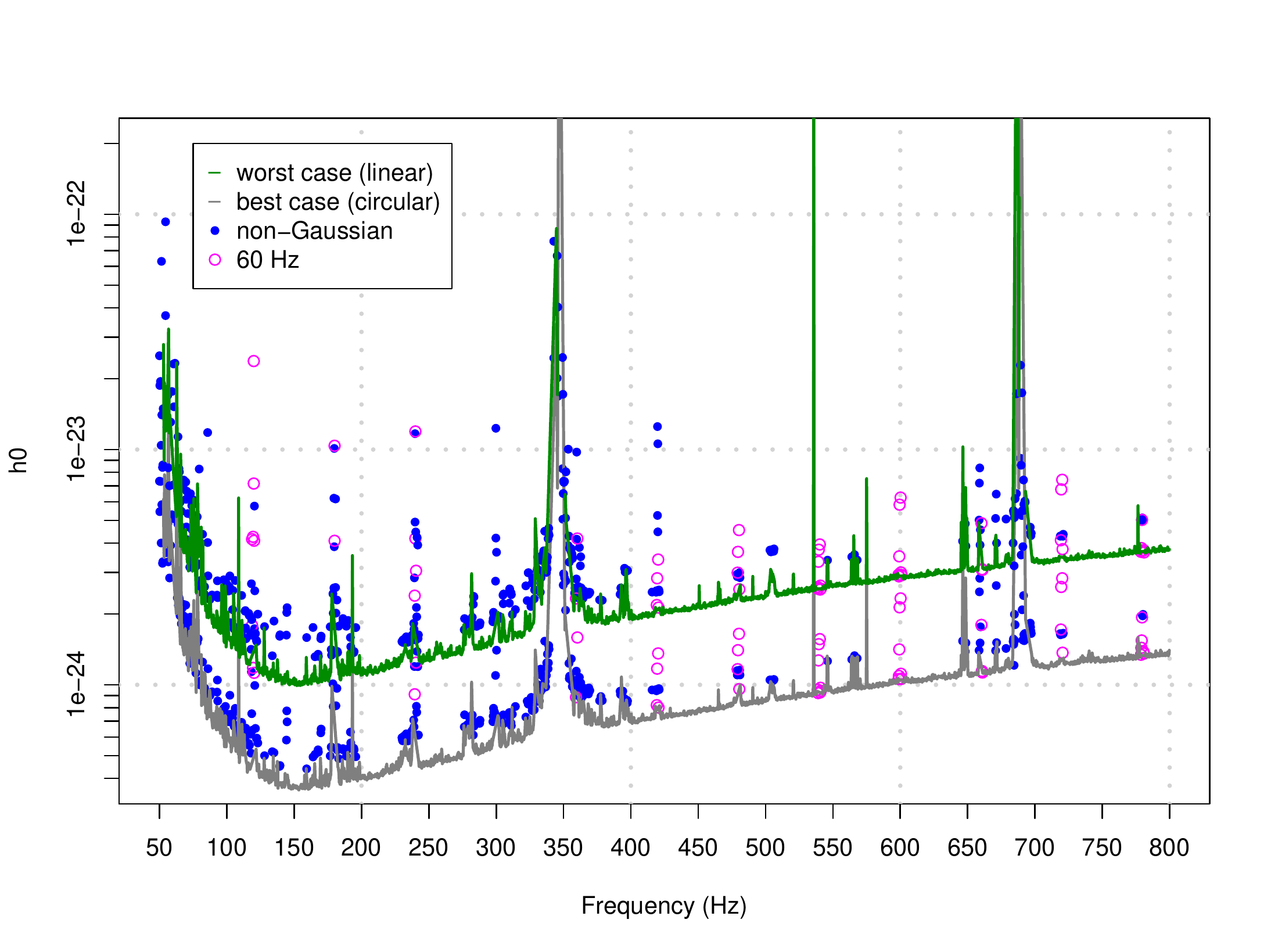}

 \caption{Full S5 upper limits. The upper (green) curve shows worst case upper limits in analyzed 0.25~Hz bands (see Table \ref{tab:excluded_ul_bands} for list of excluded bands). The lower (grey) curve shows upper limits assuming circularly polarized source. The values of solid points (marking non-Gaussian behaviour) and circles (marking power line harmonics) are not considered reliable. They are shown to indicate contaminated bands. (color online)}
\label{fig:full_s5_upper_limits}
\end{center}
\end{figure*}

The 95\% confidence level upper limits (see Fig.~\ref{fig:full_s5_upper_limits}) produced in the first stage are based on the overall noise level and largest outlier in strain found for every template in each 0.25~Hz band in
the first stage of the pipeline.
A followup search for detection is carried out for high-SNR outliers found in the first stage. 
An important distinction is that we do not report upper limits for certain frequency ranges because of contamination by detector artifacts and thus unknown statistical properties. However, the detection search used all analyzed frequency bands with reduced sensitivity in contaminated regions.

From the point of view of the analysis code the contamination by detector artifacts can be roughly separated into regions of non-Gaussian noise statistics, 60~Hz harmonics and other detector disturbances such as steeply sloped spectrum or sharp instrumental lines due to data acquisition electronics.

\section{PowerFlux algorithm and establishment of upper limits}

The data of the fifth LIGO science run was acquired over a period of nearly two years and comprised over 80000 1800-second Hann-windowed 50\%-overlapped short Fourier transforms (SFTs). Such a large dataset posed a significant challenge to the previously described PowerFlux code \cite{S4IncoherentPaper, PowerFluxTechNote, PowerFlux2TechNote}:
\begin{itemize}
 \item A $1$~Hz band (a typical analysis region) needed more than a gigabyte of memory to store the input data. 
 \item The large timebase necessitates particularly fine spindown steps of $\sci{3}{-11}$~Hz/s which, in turn requires $201$ spindown steps to cover the desired range of $[\sci{-6}{-9}, 0]$~Hz/s. The previous searches \cite{S4IncoherentPaper, EarlyS5Paper} had iterated over only 11 spindown values.
 \item The more sensitive data exposed previously unknown detector artifacts that required thorough study.
\end{itemize}

To overcome these issues, the PowerFlux analysis code was rewritten to be more memory efficient, to achieve a 10$\times$ reduction in large-run computing time and to provide more information 
useful in the followup detection search. Changes in architecture allowed us to implement the {\em Loosely Coherent} statistic \cite{loosely_coherent} which was invaluable in automating the detection search and pushing down the outlier noise floor. This is discussed in more detail in section \ref{sec:followup}.

\begin{figure}[htbp]
\begin{center}
  \includegraphics[width=3.0in]{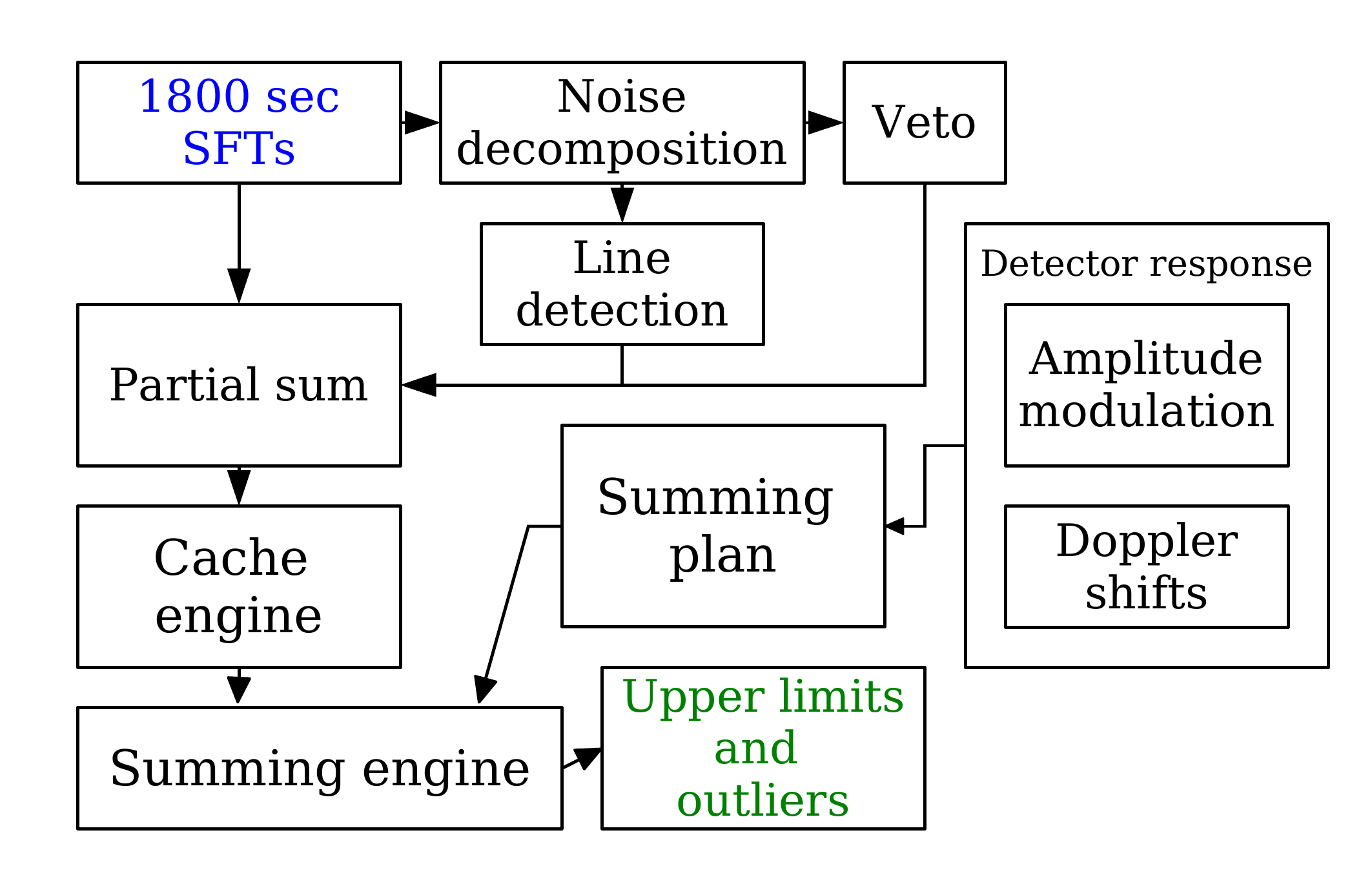}

 \caption{Flowchart of PowerFlux code (color online).}
\label{fig:PowerFlux2Flowchart}
\end{center}
\end{figure}

A flowchart of the PowerFlux program is shown in Fig.~\ref{fig:PowerFlux2Flowchart}. There are three major flows of data. The detector response involves computation of amplitude response, detector position and Doppler shifts based on knowledge of sky location searched and timing of the input data. The data set is characterized by computing data quality statistics independent of sky position. Finally, the weighted power sums are computed from the input data, folding in information on detector response and data quality to optimize performance of the code that searches over all sky positions, establishes upper limits and finds outliers.

The noise decomposition, instrumental line detection, SFT veto and detector response components are the same as in the previous version of PowerFlux.

The power sum code has been reworked to incorporate the following improvements:
\begin{itemize}
 \item 
Instead of computing power sums for specific polarizations for the
 entire dataset, we compute {\it partial power sums}: terms in the
 polarization response that are additive functions of the data.
  This allows us to sample more polarizations, or to combine or omit
   subsets of data, at a small penalty in computing cost.

\item The partial power sums are cached, greatly reducing redundant computations.
 \item The partial power sums are added hierarchically (see \ref{SFT_partitioning}) by a summing engine which makes it possible to produce simultaneously upper limits and outliers for different combinations of interferometers and time segments. This improvement significantly reduces the time needed for the followup analysis and makes possible detection of long duration signals present in only part of the data.
 \item 
Instead of including the frequency evolution model in the summing
   engine, the engine takes a {\it summing plan} (representing a
   series of frequency shifts), and contains heuristics to improve
   cache performance by partitioning SFTs based on the summing plan.
   A separate module generates the summing plan for a specific
   frequency evolution model.  This will allow us in the future to add
   different frequency models while still using the same caching and
   summing code.
\end{itemize}

\subsection{Input data}
The input data to our analysis is a sequence of $1800$-second short Fourier transforms (SFTs) which we view as a matrix $\left\|z_{t, f} \right\|$. Here $t$ is the GPS time of the start of a short Fourier transform, while $f$ denotes the frequency bin in units of $1/1800$ Hz. The SFTs are produced from
the calibrated gravitational strain channel $h(t)$ sampled at 16384~Hz.

This data is subjected to noise decomposition \cite{S4IncoherentPaper, PowerFluxTechNote} to determine the noise levels in individual SFTs and identify sharp instrumental lines. The noise level $n_t$ assigned to each SFT is used to compute SFT weight as inverse square $1/n_t^2$.

Individual SFTs with high noise levels or large spikes in the underlying data are then removed from the analysis. For a typical well-behaved frequency band, we can exclude 8\% of the SFTs while losing only 4\% of accumulated weight.
For a band with large detector artifacts (such as instrumental lines arising from resonant
vibration of mirror suspension wires), however, we can end up removing most, if not all, SFTs. As such bands are not expected to have any sensitivity of physical interest they were excluded from the upper limit analysis (Table \ref{tab:excluded_ul_bands}).

\begin{table*}[htbp]
\begin{center}
\begin{tabular}{lp{10cm}}\hline
Category & Description \\
\hline \hline
60 Hz harmonics & Anything within 1.25~Hz of a multiple of 60~Hz \\
First harmonic of violin modes & From 323~Hz to 357~Hz  \\
Second harmonic of violin modes & From 685~Hz to 697~Hz \\
Other low frequency  & 0.25~Hz bands starting at 
50.5,
51,
52,
54, 54.25,
55,
57,
58,
58.5, 58.75, 
63,
65,
66,
69,
72,
78.5,
79.75,
80.75~Hz
 \\
Other high frequency & 0.25~Hz bands starting at 
105.25,
106,
119.25,
121,
121.5,
135.75,
237.75,
238.25, 238.5,
241.5,
362~Hz \\
\hline
\end{tabular}
\caption[Frequency regions excluded from upper limit analysis]{Frequency regions excluded from upper limit analysis. These are separated into power line harmonics, harmonics of ``violin modes'' (resonant vibrations of the wires which suspend the many mirrors of the interferometer), and a number of individual bands.}
\label{tab:excluded_ul_bands}
\end{center}
\end{table*}

\subsection{PowerFlux weighted sum}
PowerFlux detects signals by summing power in individual SFTs weighted according to the noise levels of the individual SFTs and the 
time-dependent amplitude responses:
\begin{equation}
P[f_t, a_t]=\frac{\sum_{t\in SFTs} |z_{t, f_t}|^2 |a_{t}|^2 /n_t^4 }{ \sum_{t\in SFTs} |a_{t}|^4/n_t^4 }\ed
\end{equation}
Here we use $a_t$ for the series of amplitude response coefficients for a particular polarization and direction on the sky, 
$f_t$ denotes the series of frequency bin shifts due to Doppler effect and spindown, and $|z_{t, f_t}|^2$ is the power in bin 
$f_t$ from the SFT acquired at time $t$. The values $n_t$ describe levels of noise in individual SFTs and do not depend on sky location or polarization. 

The frequency shifts $f_t$ are computed according to the formula
\begin{equation}
\label{eqn:source_evolution}
f_t=f_\textrm{source}+\fdot(t-t_{\textrm{ref}})+f_\textrm{source} \frac{\vec{e}\cdot\vec{v}_t}{c}+\delta f\ec
\end{equation}
where $\fdot$ is the spindown, $f_\textrm{source}$ is the source frequency, $\vec{e}$ is the unit vector pointing toward the sky location of interest, and $\vec{v}_t$ is the precomputed detector velocity at time $t$. The offset $\delta f$ is used to sample frequencies with resolution below the resolution of a single SFT. This approximate form separating contributions from Doppler shift and spindown ignores negligible second order terms.

For each sky location, spindown, and polarization, we
compute the statistic $P[f_t+k\Delta f,a_t]$ at 501 frequencies
separated by the SFT bin size $\Delta f=1/1800$~Hz. The historical reason for using this particular number of frequency bins is that it is large enough to yield reliable statistics while small enough that a large fraction of frequency bands avoids the frequency comb of 1-Hz harmonics that emerge in long integration of the S4 data and arise from
the data acquisition and control electronics. The relatively large stepping in frequency makes the statistical distribution of the entire set stable against changes in sky location
and offsets in frequency. To obtain sub-bin resolution the initial frequency $f_t$ can be additionally shifted by a fraction of the SFT frequency bin. The number of sub-bin steps - ``frequency zoom factor'' - is documented in table \ref{tab:coincidence_parameters}.

Except at very low frequencies (which are best analyzed using methods that take phase into account), the amplitude modulation coefficients respond much more slowly to change in sky location than do frequency shifts. Thus the spacing of sky and spindown templates is determined from the behaviour of the series $f_t$. The spindown spacing depends on the inverse of the timebase spanned
   by the entire SFT set.  The sky template spacing depends on the
   Doppler shift, which has two main components: the Earth's rotation,
   which contributes a component on the order of
   $\sci{1}{-6}f_\textrm{source}$ with a period of one sidereal day; and the
   Earth's orbital velocity, which contributes a larger component of
   $~\sci{1}{-4}f_\textrm{source}$ but with a longer annual period.

If not for the Earth's rotation, all the evolution components would have evolved slowly compared to the length of the analysis and the computation could proceed by subdividing the entire dataset into shorter pieces which could be sampled on a coarser grid and then combined using finer steps. We can achieve a very similar result by grouping SFTs within each piece by (sidereal) time of day, which has the effect of freezing the Earth's rotation within each group.

A further speedup can be obtained by reduction in template density, which is allowed by degeneracy between contributions from spindown mismatch and orbital velocity shift arising from mismatch in sky location.

\subsection{Partial power sum cache}
The optimizations just described can all be made simultaneously by implementing an associative cache of previously computed power sums. This approach also has the advantage of being able to accommodate new frequency evolution models (such as emission from a binary system) with few modifications.

The cache is constructed as follows. First, we subdivide the sky into {\em patches} small enough that amplitude response coefficients  can be assumed constant on each patch. Each set of templates from a single patch is computed independently using amplitude response coefficients from a representative template of its patch.

Second, we separate the weighted power sum into the numerator and denominator sums:
\begin{equation}
\label{eqn:PS}
PS[f_{t}, b_{t}]=\sum_{t\in \textrm{SFTs}}  \frac{b_{t}}{n_t^4} |z_{t, f_{t}}|^2
\end{equation}
\begin{equation}
\label{eqn:WS}
WS[c_{t}]=\sum_{t\in \textrm{SFTs}}  \frac{c_{t}}{n_t^4}\ec
\end{equation}
where values for a fixed set of amplitude response coefficients $b_t$ and $c_t$ (discussed in the next section) are stored in the partial power sum cache with the frequency shift series $f_t$ used as a key. 
The fact that both sums are additive functions of the set of SFTs for which they are computed allows partial power sums to be broken into several components and then recombined later.

\subsection{Polarization decomposition}

While it is efficient to compute the partial power sums for a small number of polarizations, one can also decompose the coefficients $b_t$ and $c_t$ into products of detector-specific time-dependent parts and static coefficients that depend on polarization alone. This analysis extends \cite{PowerFluxPolarizationNote, gen_powerflux}.

First, we introduce quadratic and quartic detector response series:
\begin{equation}
\label{eqn:f2}
F^{2, i}_t=  (F^+_t)^{2-i} (F^\times_t)^i
\end{equation}
\begin{equation}
\label{eqn:f4}
F^{4, i}_t=  (F^+_t)^{4-i} (F^\times_t)^i
\end{equation}
(with $i=0-2$ and $i=0-4$, respectively), 
and the corresponding sets of polarization response coefficients:
\begin{equation}
\label{eqn:a2}
\begin{array}{l}
A^{2, 0}(\iota, \psi)=\frac{1}{8}(1+\cos^2(\iota))^2(1+\cos(4\psi))+\\
\quad\quad\quad\quad + \frac{1}{4}\cos^2(\iota)(1-\cos(4\psi)) \\
A^{2, 1}(\iota, \psi)=\left(\frac{1}{4}(1+\cos^2(\iota))^2-\frac{1}{2}\cos^2(\iota)\right)\sin(4\psi) \\
A^{2, 2}(\iota, \psi)=\frac{1}{8}(1+\cos^2(\iota))^2(1-\cos(4\psi))+\\
\quad\quad\quad\quad +\frac{1}{4}\cos^2(\iota)(1+\cos(4\psi)) \\
\end{array}
\end{equation}
\begin{equation}
\label{eqn:a4}
\begin{array}{l}
A^{4, 0}=A^{2,0}A^{2,0} \\
A^{4, 1}=2 A^{2,0}A^{2, 1} \\ 
A^{4, 2}=2A^{2,0}A^{2, 2}+A^{2,1}A^{2,1} \\
A^{4, 3}=2A^{2,2}A^{2,1} \\
A^{4, 4}=A^{2,2}A^{2,2} \\
\end{array}
\end{equation}
Here $\iota$ and $\psi$ are the usual \cite{jks} inclination and orientation parameters of the source.

The amplitude response coefficients can be represented as
\begin{equation}
\begin{array}{l}
b_t(\iota, \psi)=\sum_{i=0}^2 F_t^{2, i} A^{2, i}(\iota, \psi)\\
c_t(\iota, \psi)=\sum_{i=0}^4 F_t^{4, i} A^{4, i}(\iota, \psi)
\end{array}
\end{equation}
and, given previously computed partial power sums, we compute the weighted power sum for an arbitrary polarization as
\begin{equation}
P[f_{t}, \iota, \psi]=\frac{\sum_{i=0}^{2}  PS[f_{t}, F_{t}^{2, i}] A^{2, i}(\iota, \psi) } {
  \sum_{i=0}^{4}  WS[f_{t},  F_{t}^{4, i}] A^{4, i}(\iota, \psi)
   }\ed
\end{equation}
In this approach we use equations \ref{eqn:PS} and \ref{eqn:WS} to compute power sums for a non-physical but computationally convenient set of polarizations that can be combined into physical power sums in the end.

\subsection{SFT set partitioning}
\label{SFT_partitioning}
The PowerFlux weighted power sum is additive with respect to the set of SFTs it is computed with. This can be used to improve the efficiency of the cache engine, which will have a higher hit ratio for more tightly grouped SFTs. This needs to be balanced against the larger overhead from accumulating individual groups into the final weighted power sum. In addition, larger groupings could be used to analyze subsegments of the entire run, with the aim of detecting signals that were present only during a portion of the 2 years of  data.

In this analysis, we have used the following summing plan:
First, for each individual detector the SFT set is broken down into equally spaced chunks in time. Five chunks per detector were used in the analysis of the low frequency range of 50-400~Hz, for which detector non-stationarity was more pronounced. Three chunks per detector were used for analysis of the 375-800~Hz range. 

The partial power sums for each chunk are computed in steps of 10 days each, which are also broken down into 12 groups by the magnitude of their frequency shifts. 

The individual groups have their frequency shift series rounded to the nearest integer frequency bin, and the result is passed to the associative cache.

\subsection{Computation of upper limits, outliers and other statistics}
Having computed partial power sums for individual chunks, we combine them into contiguous sequences, both separately by detector and as a whole, to form weighted power sums. These sums are used to establish upper limits based on the Feldman-Cousins \cite{FeldmanCousins} statistic, to obtain the signal-to-noise ratios and auxiliary statistics used for detector characterization and to assess the Gaussianity of underlying data.

An important caveat is that the sensitivity of the detectors improved considerably toward the end of the data taking run, especially at low frequencies. As the SFT weight veto described earlier is performed for the entire dataset, it can remove a considerable fraction of data from the first few chunks. Thus at frequencies below $400$~Hz, the upper limit chosen for each  
   frequency bin is the value obtained from analyzing the entire run,
   the last $4/5$ of the run, or the last $3/5$ of the run, whichever
   value is lowest.  At frequencies above $400$~Hz we use the value
   obtained from the entire run or the last $2/3$ of the run, whichever
   value is lowest.

The detection search was performed on outliers from any contiguous combination of the chunks, but we have not run tests to estimate pipeline efficiency on smaller subsets.
\subsection{Injections and Validation}

The analysis presented here has undergone extensive checking, including independent internal review of the code and numerous Monte-Carlo injection runs. We present a small portion of this work to assure the reader that the pipeline works as described.

\begin{figure}[htbp]
\begin{center}
  \includegraphics[width=3.0in]{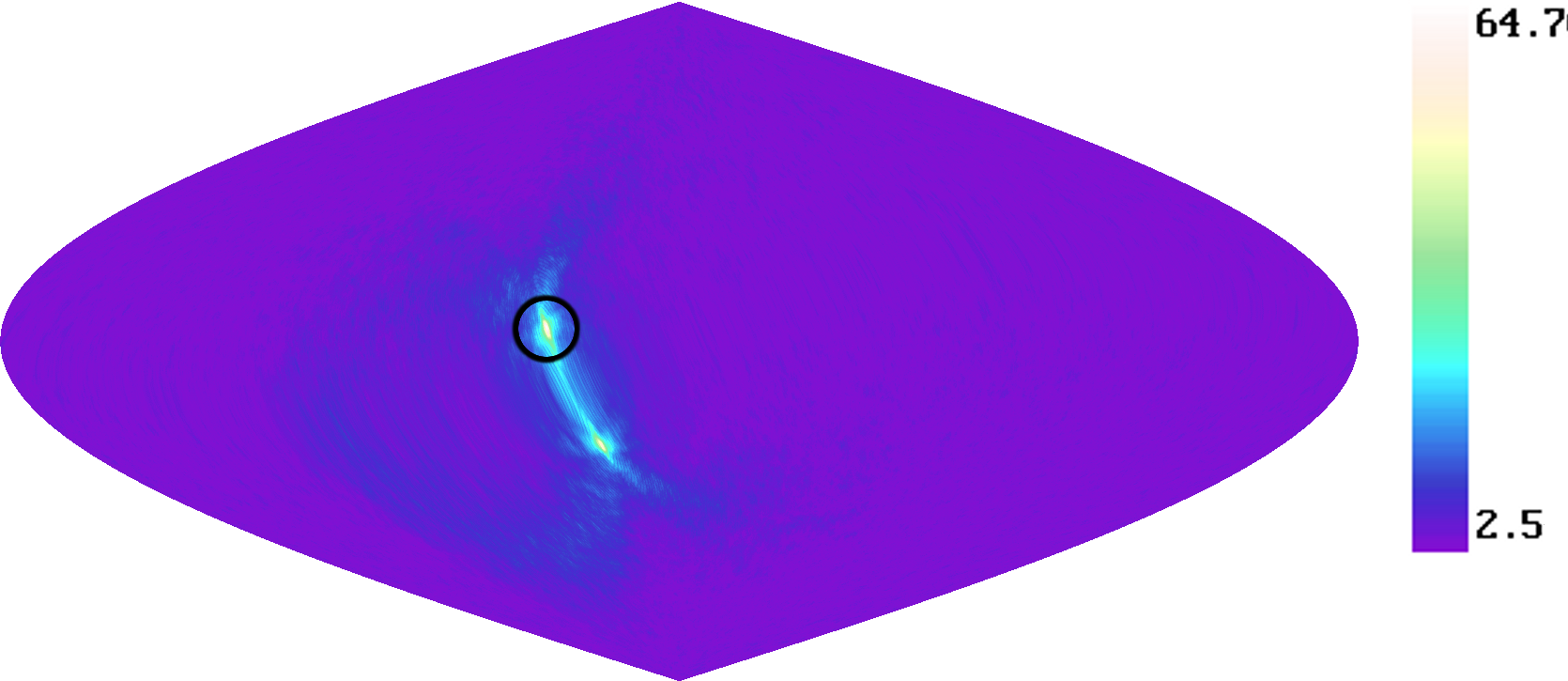}
 \caption[SNR skymap]{SNR skymap for hardware injection of a simulated signal. The circle is centered on the location of the injection. The high frequency of the signal (575~Hz) allows good localization (color online).}
\label{fig:snr_skymap}
\end{center}
\end{figure}

One of the most basic tests is correct reconstruction of hardware and software signal injections. Figure \ref{fig:snr_skymap} shows a skymap of the signal-to-noise ratio on the sky for a sample injection, for which the maximum is found at a grid point near the injection location. As the computation of weighted sums is a fairly simple algebraic transformation, one can infer the essential correctness of the code in the general case from the correctness of the skymaps for several injections.

A Monte-Carlo injection run also provides test of realistically distributed software paths, validation of upper limits and characterization of parameter reconstruction.

\begin{figure}[htbp]
\begin{center}
  \includegraphics[width=3.0in]{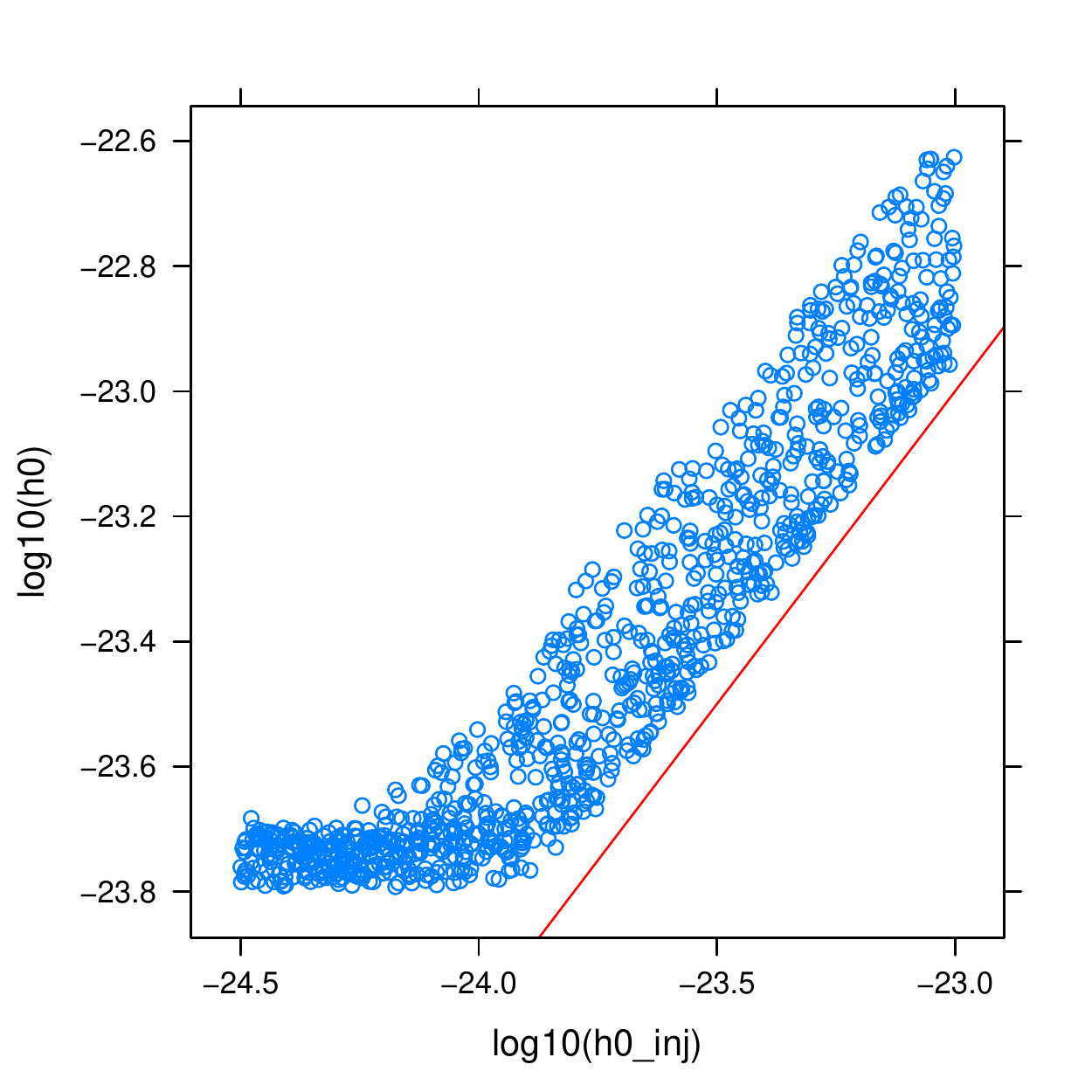}
 \caption[Upper limit versus injected strain]{Upper limit validation. Each point represents a separate injection. Established upper limit (y axis) is compared against injected strain value (x axis, red line) (color online).}
\label{fig:ul_vs_strain}
\end{center}
\end{figure}

\begin{figure}[htbp]
\begin{center}
  \includegraphics[width=3.0in]{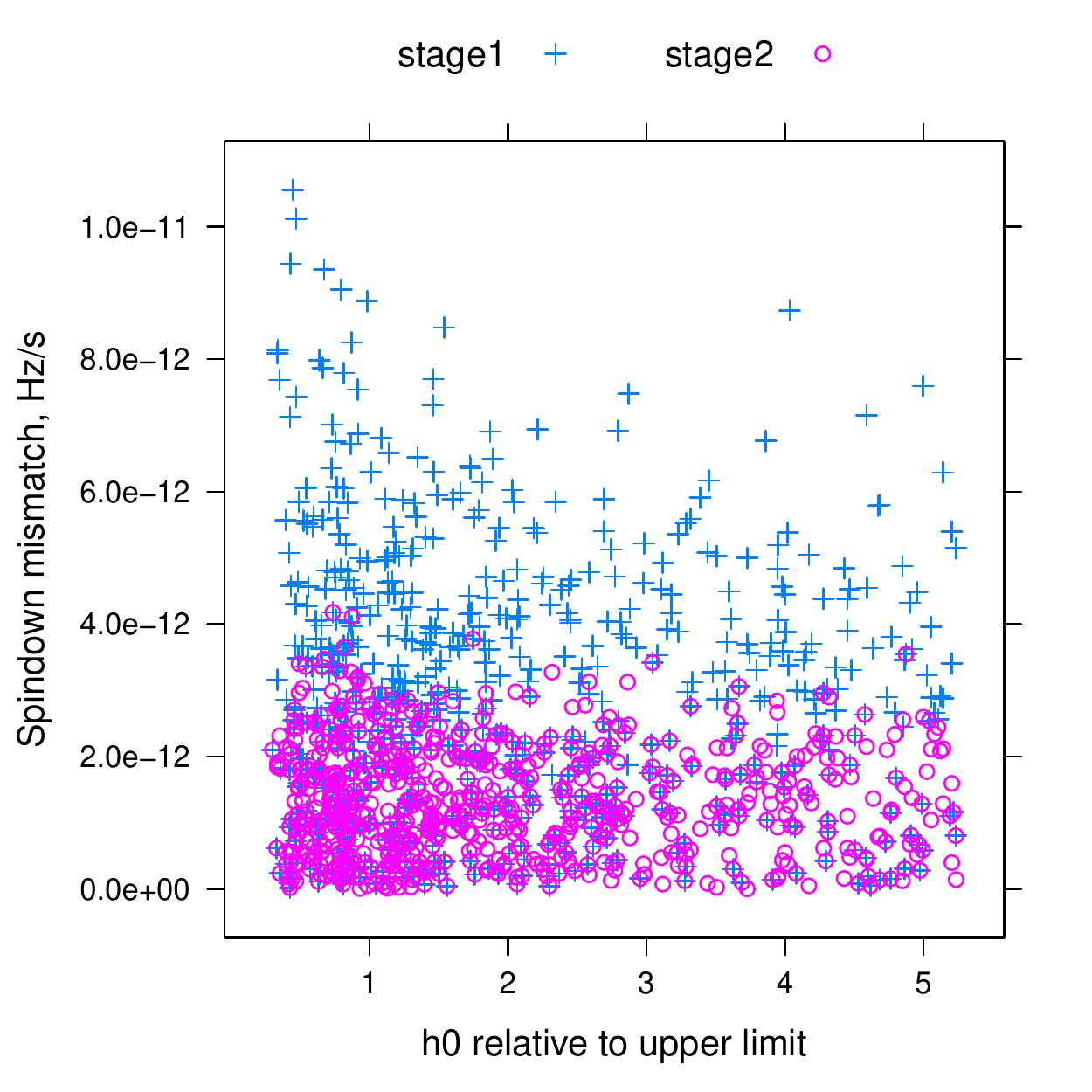}
 \caption[Spindown improvement]{Improvement of spindown localization of injected test signals. The injected strain divided by the upper limit in this band (before injection) is shown on the x axis. The difference between injection spindown and spindown of corresponding outlier is shown on the y axis. Crosses - semi-coherent, circles - loosely coherent (color online).}
\label{fig:spindown_improvement}
\end{center}
\end{figure}

\begin{figure}[htbp]
\begin{center}
  \includegraphics[width=3.0in]{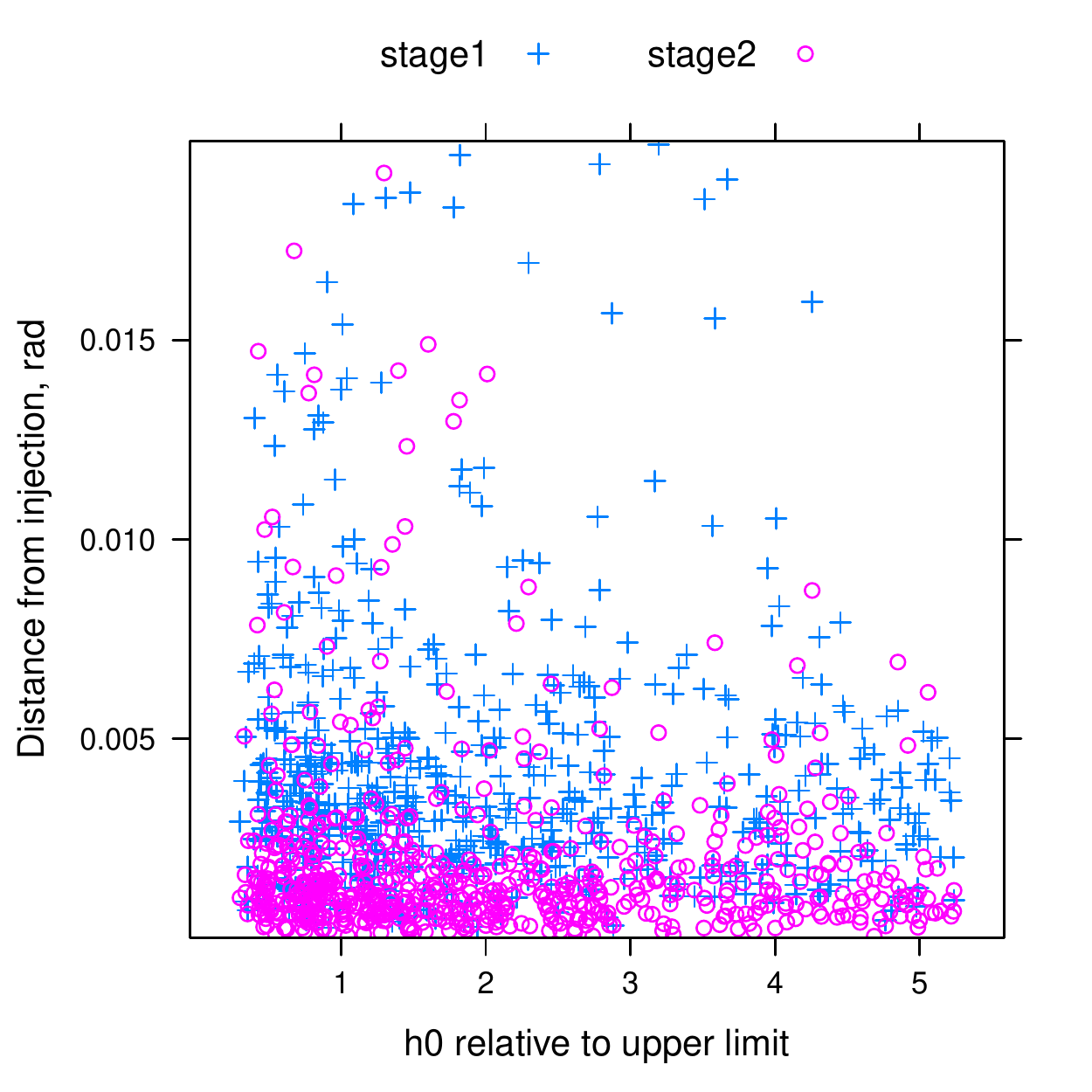}
 \caption[Distance improvement]{Improvement of position localization of injected test signals. The injected strain divided by the upper limit in this band (before injection) is shown on the x axis. The distance between injection sky location and that of corresponding outlier is shown on the y axis. Crosses - semi-coherent, circles - loosely coherent (color online).}
\label{fig:distance_improvement}
\end{center}
\end{figure}

\begin{figure}[htbp]
\begin{center}
  \includegraphics[width=3.0in]{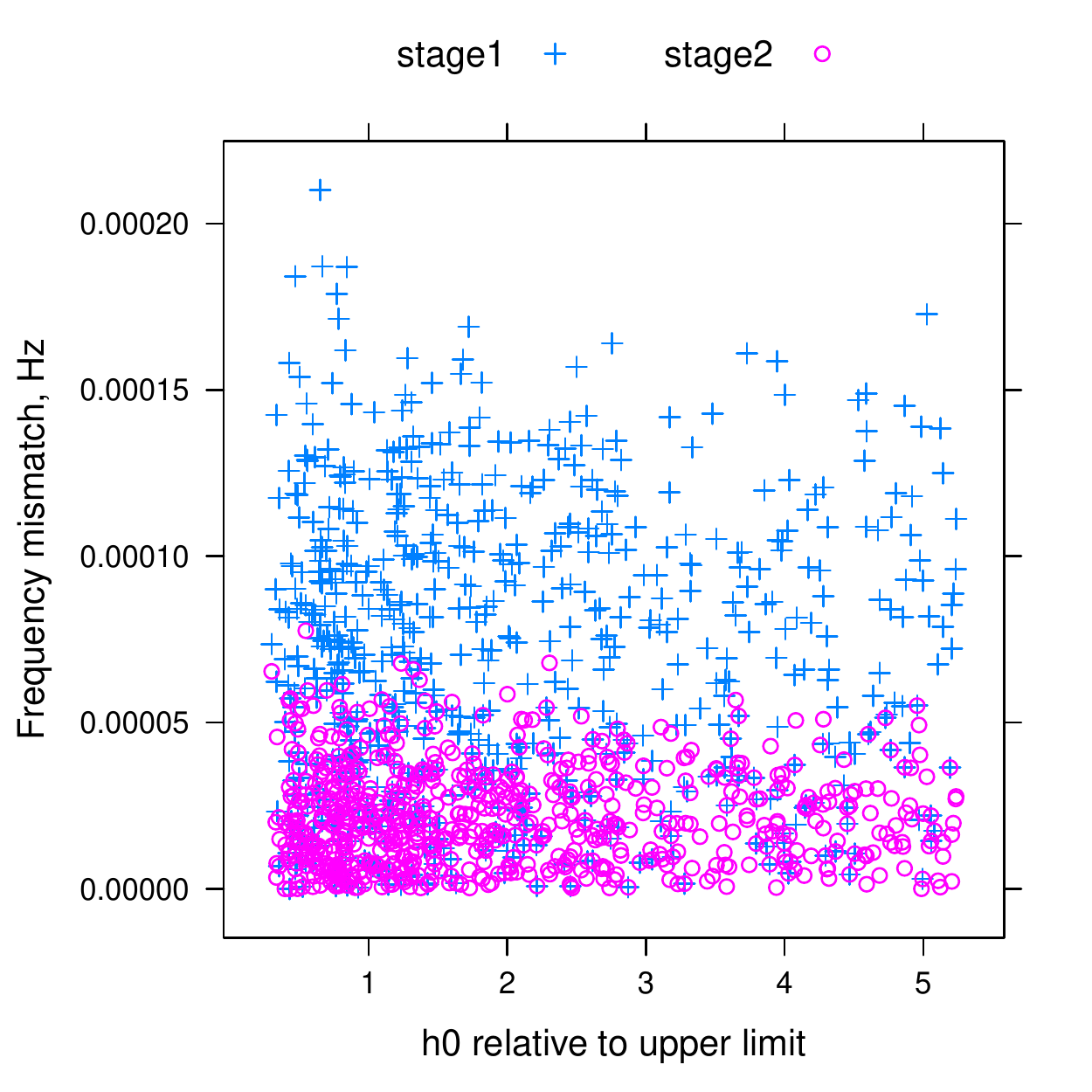}
 \caption[Frequency improvement]{Improvement of frequency localization of injected test signals. The injected strain divided by the upper limit in this band (before injection) is shown on the x axis. The difference between injection frequency and frequency of corresponding outlier is shown on the y axis. Crosses - semi-coherent, circles - loosely coherent (color online).}
\label{fig:f0_improvement}
\end{center}
\end{figure}

\begin{figure}[htbp]
\begin{center}
  \includegraphics[width=3.0in]{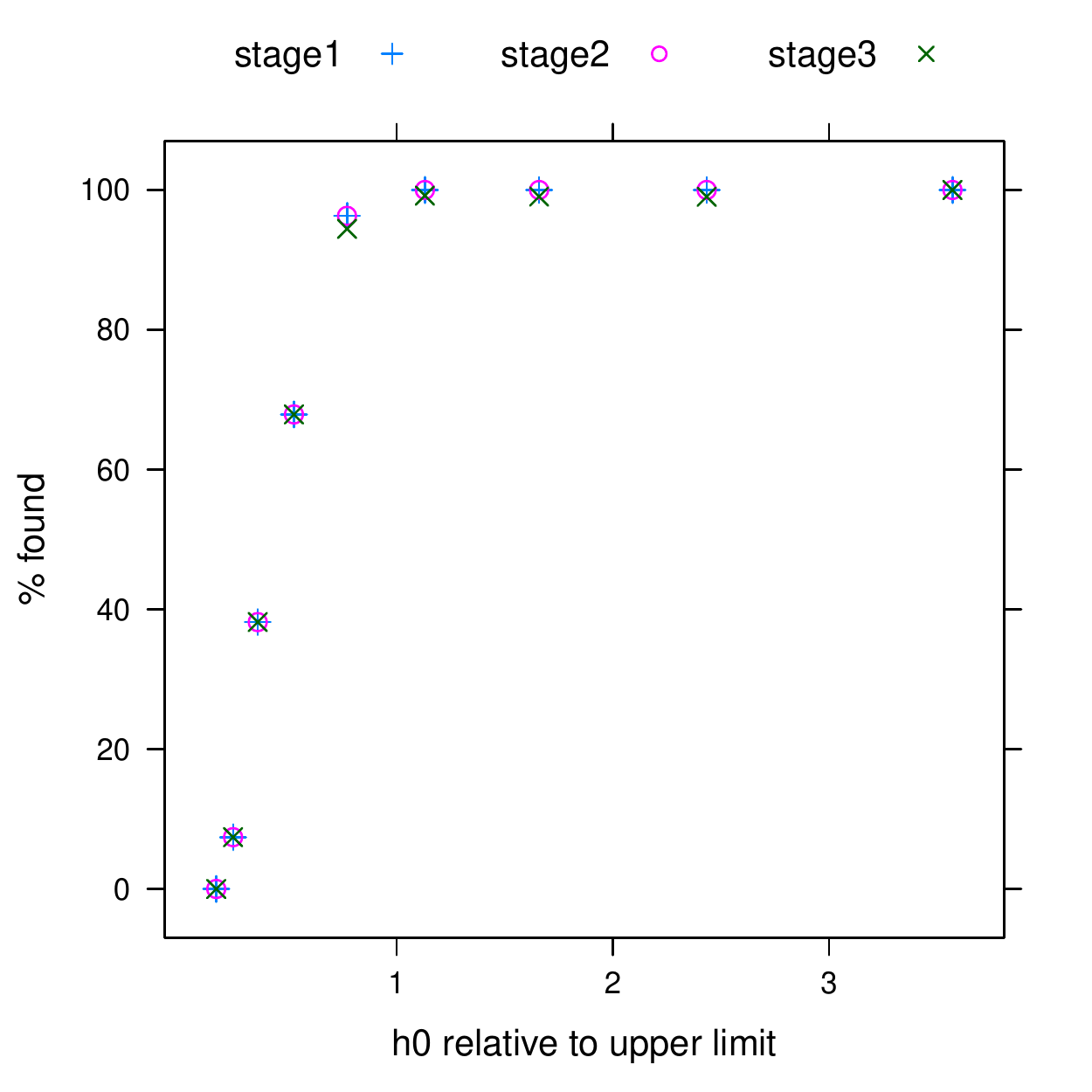}
 \caption[Injection recovery]{Injection recovery from semi-coherent analysis stage (crosses), after the first loosely coherent followup (circles) and after second stage of loosely coherent followup (diagonal crosses). The injected strain divided by the upper limit in this band (before injection) is shown on the x axis. The percentage of surviving injections is shown on the y axis (color online).}
\label{fig:injection_recovery}
\end{center}
\end{figure}

In a particular injection run we are concerned with three main issues:
\begin{itemize}
 \item The upper limits established by the search should be above injected values. Figure \ref{fig:ul_vs_strain} shows results of such a simulation at 400~Hz, confirming validity of the search.
 \item We need to determine the maximum mismatches in signal parameters the search can tolerate while still producing correct upper limits and recovering injections. Figures \ref{fig:spindown_improvement}, \ref{fig:distance_improvement},  \ref{fig:f0_improvement} show results of such analysis in the 400~Hz band. The signal localization is within the bounds used by the followup procedure (discussed in section \ref{sec:followup}).
 \item The efficiency ratio of injection recovery should be high. As seen in Fig.~\ref{fig:injection_recovery} our recovery ratio for semi-coherent search is nearly 100\% for injections at the upper limit level.
\end{itemize}

\section{Loosely Coherent code and detection pipeline}
\label{sec:followup}
The reduced sensitivity of a semicoherent method like PowerFlux relative to a fully coherent search comes with robustness to variation in phase of the input signal, be it from small perturbations of the source due to a companion or from imperfections in the detector. 

One way to achieve higher sensitivity while preserving robustness to variations in the phase of the input signal is to use a {\em Loosely Coherent} search code that is sensitive to families of signals following a specific phase evolution pattern, while allowing for fairly large deviations from it. We have extended PowerFlux with a program that computes a loosely coherent power sum. The results of simulations of this program on Gaussian noise were first presented in \cite{loosely_coherent}.

Searches for continuous-wave signals have typically been performed
using combinations of coherent and semicoherent methods.  A
coherent method requires a precise phase match between the signal  
and a model template over the entire duration of the signal, and 
thus requires a close match between the signal and model
parameters: each model template covers only a small region of the
signal space.  A semicoherent method requires phase matching over
short segments of the data and discards phase information
between segments, and each template therefore covers a larger
region of the signal space.  We use the term {\em Loosely Coherent} to 
describe a broader class of algorithms whose templates cover some
arbitrarily specified region of the signal space, with sensitivity
falling off outside of the template boundaries.
  
One way to implement a loosely coherent search 
is by requiring the signal to match a phase model very closely over
a narrow time window, but then smoothly downgrading the phase-match
requirement over longer timespans by means of a weighting kernel.
The mathematical expression of this is given in equation \ref{eqn:loosely_coherent}, below.
The allowable phase drift, expressed as radians per unit time, is a
tunable parameter of the search.  Larger allowed phase drifts  
result in templates that cover a larger region of the signal space,
but with less power to discriminate true signals from noise.

%

The variant of the loosely coherent statistic used in this paper is derived from the PowerFlux code base and is meant for analysis with wide phase evolution tolerance. It is not the most computationally efficient, but has well-understood robustness properties and suffices for followup of small sky areas. A dedicated program for future searches is under development. The technical description of the present implementation can be found in \cite{PowerFlux2TechNote}.

\subsection{Loosely coherent weighted sum}
The loosely coherent statistic is based on the same power sum computation used in the PowerFlux computing infrastructure, but instead of a single sum over SFTs, we have a double sum:
\begin{equation}
\label{eqn:loosely_coherent}
\begin{array}{l}

P[f_t, \phi_t, a_t, f_{t'}, \phi_{t'}, a_t', \delta]=\\
\quad\displaystyle\frac{\sum_{t, t'\in SFTs} e^{i\phi_{t'}-i\phi_t}K_\delta(|t-t'|) \bar{z}_{t', f_{t'}}z_{t, f_{t}} \bar{a}_t a_{t'}  /n_t^2n_{t'}^2 }{ \sum_{t, t'\in SFTs} K_\delta(|t-t'|)|a_{t}|^2|a_{t'}|^2/n_t^2n_{t'}^2 }\ed
\end{array}
\end{equation}
Here $\phi_t$ is the series of phase corrections needed to transition the data into the solar system barycenter frame of reference and to account for source evolution between times $t$ and $t'$. 

The formula \ref{eqn:loosely_coherent} is generic for any second order statistic, a nice description is presented in \cite{cross_correllation} as generalization of cross-correlation.  

In order to make a statistic \ref{eqn:loosely_coherent} loosely coherent we need to make sure that it admits signals with phase deviation up to a required tolerance level and rejects signals outside of that tolerance. We achieve this by selecting a low pass filter for the kernel $K_\delta(|t-t'|)$. A $\sinc=\frac{\sin(x)}{x}$ based filter provides the steepest rejection of signals with large phase deviation, but is computationally expensive. Instead, we use the Lanczos kernel with parameter 3:
\begin{equation}
\tilde{K}_\delta(\Delta t)=\left\{
\begin{array}{ll}
\sinc\left(\frac{\delta \Delta t}{0.5\textrm{hr}}\right)\sinc\left(\frac{\delta \Delta t}{1.5\textrm{hr}}\right)& \textrm{when~} \frac{\delta \Delta t}{0.5\textrm{hr}} < 3\pi\\
0.0 & \textrm{otherwise}  \\
\end{array}
\right.
\end{equation}
The choice of low-pass filter implies that the terms with $t=t'$ are always included.

The parameter $\delta$ determines the amount of accumulated phase mismatch that is permitted over 30~minutes. For large values of $\delta$ the search is more tolerant to phase mismatch and closer in character to PowerFlux power sums. For smaller values the effective coherence length is longer, requiring finer template spacing and yielding higher signal-to-noise ratios.

\subsection{Partial power sum cache}
The partial power cache is constructed similarly to the PowerFlux case. Instead of a single series of frequency shifts $f_t$ the partial power sums depend on both $f_t$ and $f_{t'}$. The additional key component consists of a series of differential Doppler shifts, from the derivative of Doppler shift with respect to frequency, since a small change in frequency has a large effect on phases $\phi_t$ and $\phi_{t'}$.

For the value of $\delta=\pi/2$ used in this analysis the cross terms ($t\neq t'$) are often zero as the Lanczos kernel vanishes for widely separated SFTs. The SFT partitioning scheme takes advantage of this by forming smaller groups and only computing cross terms between groups that are close enough to produce non-zero results. 


\subsection{Polarization decomposition}
The polarization decomposition for the loosely coherent search is similar to that used by PowerFlux. The two changes required are the treatments of coefficients involving both cross and plus detector response terms and imaginary terms \cite{PowerFlux2TechNote}.

The implementation used in this search is obtained by the mathematical method of polarization\footnote{This has nothing to do with polarization of gravitational wave signals, but refers to the fact that the map between symmetric multilinear forms $L(x, y, z, ...)$ with $k$ arguments and homogeneous polynomials $P$ of degree $k$ given by $P(x)=L(x, x, x, ...)$ is a bijection. } of homogeneous polynomials of equations \ref{eqn:f2} and \ref{eqn:f4}:

\begin{equation}
\begin{array}{l}
F^{2, 0}(t, t')=F^+_t F^+_{t'} \\
F^{2, 1}(t, t')=\frac{1}{2}\left(F^+_t F^\times_{t'}+F^\times_t F^+_{t'}\right) \\
F^{2, 2}(t, t')=F^\times_t F^\times_{t'} \\
\end{array}
\end{equation}

\begin{equation}
\begin{array}{l}
F^{4, 0}=F^{2,0}F^{2,0} \\
F^{4, 1}=F^{2,0}F^{2, 1} \\ 
F^{4, 2}=\frac{1}{3}F^{2,0}F^{2, 2}+\frac{2}{3}F^{2,1}F^{2,1} \\
F^{4, 3}=F^{2,2}F^{2,1} \\
F^{4, 4}=F^{2,2}F^{2,2} \\
\end{array}
\end{equation}
Equations \ref{eqn:f2} and \ref{eqn:f4} are obtained by setting $t'=t$.
This allows us to compute the real part of the product $\bar{a}_{t'}a_{t}$ using the same polarization coefficients $A^{2,i}$ and $A^{4, i}$ as used by the PowerFlux search:
\begin{equation}
\Re (\bar{a}_{t'}a_{t})=\sum_{i=0}^{2} F^{2, i} A^{2, i}\ed
\end{equation}
The imaginary part is equal to
\begin{equation}
\Im (\bar{a}_{t'}a_{t})=\frac{1}{8}\left(F^+_t F^\times_{t'}-F^\times_t F^+_{t'}\right)(1+\cos^2\iota)\cos \iota
\end{equation}
and is neglected in the analysis. This approximation is justified for several reasons. First, the difference $F^+_t F^\times_{t'}-F^\times_t F^+_{t'}$ is small relative to other terms for closely spaced $t$ and $t'$. Second, the part depending on $\cos(\iota)$ is large for polarizations close to circular, to which we are more sensitive anyway. The simulations have shown that discarding this term reduces the SNR by about 4\% for circular polarizations.

\subsection{Followup procedure}

The detection pipeline consists of three stages. The first stage is a regular PowerFlux run that produces lists of outliers with single-interferometer SNR of 5 or greater. 

The outliers are subjected to a coincidence test (parameters shown in Table \ref{tab:coincidence_parameters}), where the outliers from the multi-interferometer data with SNR of at least 7 are compared against nearby single-interferometer outliers. Frequency consistency provides the tightest constraint, with sky position and spindown helping to eliminate loud instrumental artifacts. As the ability to localize signals depends largely on Doppler shifts from Earth orbital motion we project outlier locations onto the ecliptic plane to compute ``ecliptic distance'' for a sky coincidence test. A number of 0.1-Hz regions (see Table \ref{tab:excluded_bands}) had so many coincidences (due to highly disturbed local spectra) that they had to be excluded from the analysis.

The outliers at nearby frequencies and sky locations are grouped together, and only the loudest is passed to the next stage of followup.

During the second stage the resulting outliers are analyzed using the loosely coherent code with phase mismatch parameter $\delta=\pi/2$, while combining data from different interferometers incoherently. The sky resolution is made finer (``zoomed'') by a factor of $4$. The incoherent combination provides SNR data both for individual interferometers, as well as for their combination, while being faster to compute due to fewer terms in the double sum in equation \ref{eqn:loosely_coherent}.

\begin{figure}[htbp]
\begin{center}
  \includegraphics[width=3.0in]{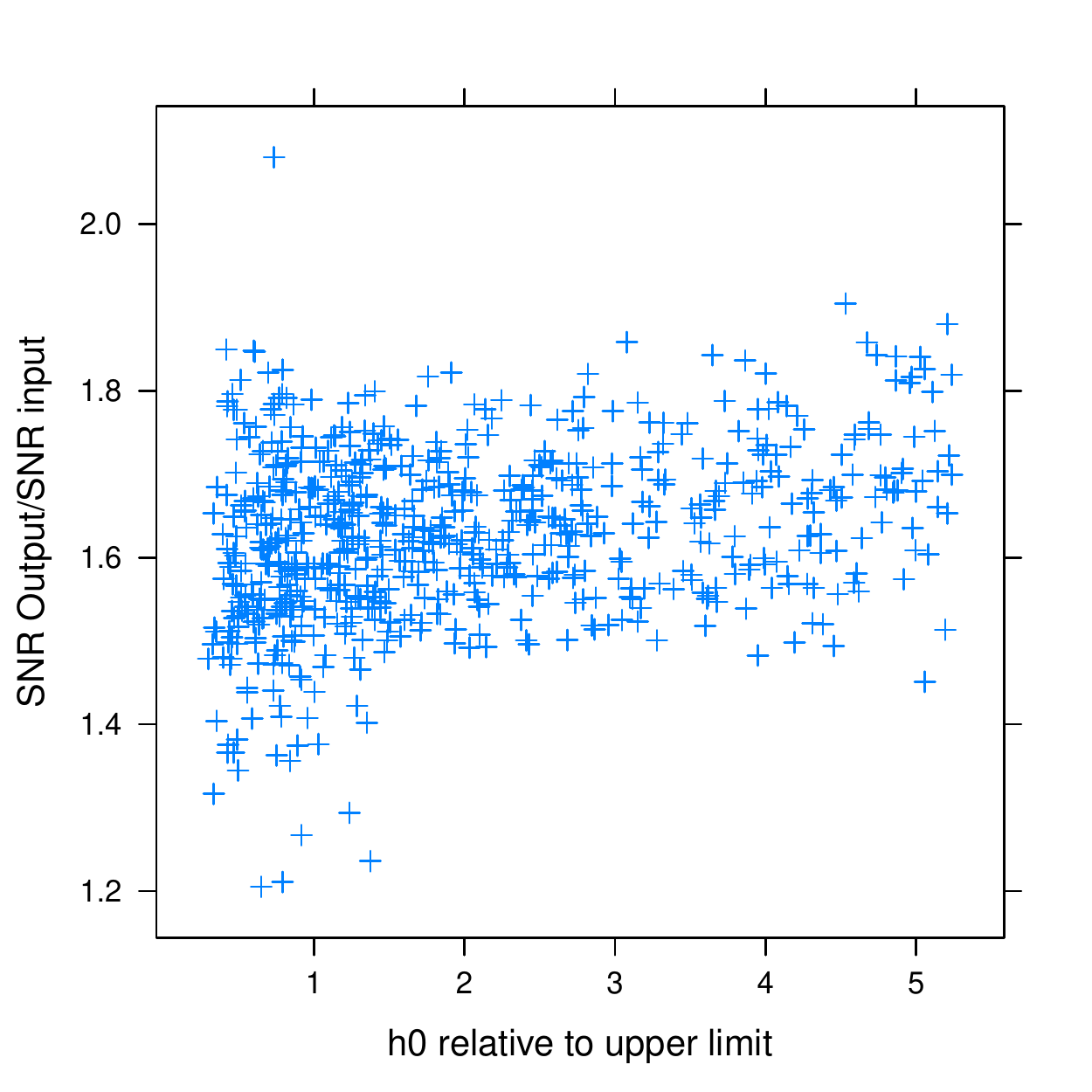}
 \caption[SNR improvement from first followup stage]{Relative increase in SNR  of injected test signals from semi-coherent to loosely coherent stage. The plot is restricted to injections that passed the coincidence test. The injected strain divided by the upper limit in this band (before injection) is shown on the x axis. The ratio of loosely coherent SNR to semi-coherent SNR is shown on the y axis (color online).}
\label{fig:snr_improvement}
\end{center}
\end{figure}

The outliers in SNR passing the loosely coherent analysis are required to show at least a 20\% increase in multi-interferometer SNR while not shifting appreciably in frequency. The required SNR increase from the semi-coherent to the loosely coherent stage is quite conservative, as can be seen in Fig.~\ref{fig:snr_improvement}. In addition, we apply a  minimum SNR cut: the SNR of each individual interferometer should be at least 20\% of the multi-interferometer SNR. This condition is essential to eliminating coincidences from loud instrumental lines in only one interferometer.  

In the third stage of followup the remaining outliers are reanalyzed with the loosely coherent pipeline, which now coherently combines data from both interferometers. To eliminate the possibility of a relative global phase offset we sampled 16 possible phase offsets between interferometers. This step is merely a precaution that was easy to implement - we did not expect to see a significant offset, as the relative interferometer timing was determined to be within $10\mu s$ \cite{LIGO_detector, S5_calibration}.

\begin{figure}[htbp]
\begin{center}
  \includegraphics[width=3.0in]{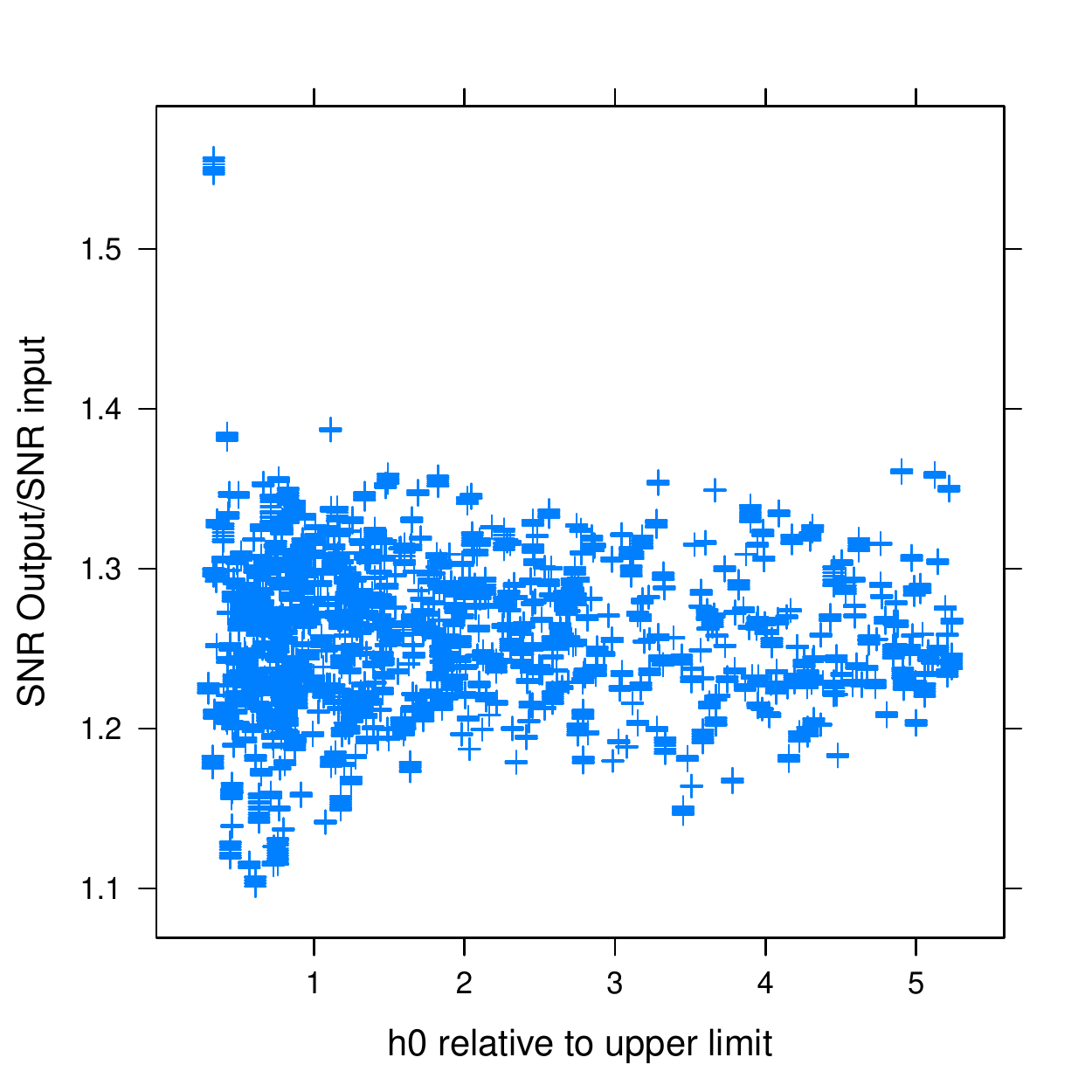}
 \caption[SNR improvement from second followup stage]{Relative increase in SNR from coherent combination of data from different interferometers. The plot is restricted to injections that passed the coincidence test. The injected strain divided by the upper limit in this band (before injection) is shown on the x axis. The ratio of coherently combined loosely coherent SNR to loosely coherent SNR from the previous stage is shown on the y axis (color online).}
\label{fig:snr_improvement2}
\end{center}
\end{figure}

Each outlier was then required to show the expected increase in SNR of at least 7\% over the value from the second stage of followup, while maintaining the same frequency tolerance. The improvement in signal-to-noise ratio seen in simulations is shown in Fig.~\ref{fig:snr_improvement2}. Most injections have greater than  10\% increase in SNR leaving room for possible mismatch in phase of up to $\pi/16$.

\subsection{Performance of the detection pipeline}

Every detection pipeline can be described by two figures of merit - false alarm ratio and recovery ratio of true signals. 

Since our analysis is computationally limited, we can use a more sensitive code to confirm or reject outliers. Thus, our main objective in optimizing each pipeline stage was to have as high a recovery ratio as possible, while generating a small enough false alarm ratio to make the subsequent step computationally feasible.

The recovery ratios found in a Monte-Carlo simulation of first, second and third followup stages are shown in Fig.~\ref{fig:injection_recovery}. The graph shows that the loosely coherent stages have less than a 5\% loss ratio of injections, and the overall pipeline performance approaches 100\% right at the upper limit threshold. 

While it is possible to compute the false alarm ratio for Gaussian noise, this number is not very informative, since most outliers are the result of instrumental artifacts, as discussed in section \ref{sec:results}.
\begin{table*}[htbp]
\begin{center}
\begin{tabular}{ccl}\hline
Center frequency (Hz) & Width (Hz) & Description \\
\hline \hline
63 & 0.1 & Pulsed heating \\
64 & 0.1 & 16 Hz harmonic from data acquisition system \\
66 & 0.1 & Pulsed heating \\
67 & 0.1 & Unidentified strong line in L1\\
69 & 0.1 & Pulsed heating \\
75 & 0.1 & Unidentified strong line in L1\\
96 & 0.1 & 16 Hz harmonic from data acquisition system \\
100 & 0.1 & Unidentified strong line in H1\\
\hline
\end{tabular}
\caption[Frequency regions excluded from coincidence test]{Frequency regions excluded from the coincidence test because of severe noise contamination leading to numerous outliers inconsistent with a true signal.}
\label{tab:excluded_bands}
\end{center}
\end{table*}

\begin{table*}[htbp]
\begin{center}
\begin{tabular}{lccc}\hline
Parameter & 50-100~Hz & 100-400~Hz & 400-800~Hz \\
\hline \hline
\multicolumn{4}{c}{\em Main run} \\
frequency zoom factor& 2 & 2 & 2 \\
sky map zoom factor& 1 & 1 & 1 \\
spindown step (Hz/s) & $\sci{3}{-11}$ & $\sci{3}{-11}$ & $\sci{3}{-11}$ \\
\multicolumn{4}{c}{\em First coincidence step} \\
maximum frequency mismatch (mHz) & $2$ & $1$ & $1$ \\
maximum ecliptic distance (radians) & $0.25$ & $0.06$ & $0.03$ \\
maximum spindown mismatch (Hz/s) & $\sci{6}{-11}$ & $\sci{2}{-11}$ & $\sci{2}{-11}$ \\
minimum multi-interferometer SNR & 7 & 7 & 7 \\
minimum single-interferometer SNR & 5 & 5 & 5 \\
\multicolumn{4}{c}{\em Loosely coherent followup} \\
phase mismatch (radians) & $\pi/2$ & $\pi/2$ & $\pi/2$ \\
followup disk radius (radians)& $0.25$ & $0.05$ & $0.03$ \\
followup spindown mismatch (Hz/s) & $\sci{2}{-11}$ & $\sci{2}{-11}$ & $\sci{2}{-11}$ \\
frequency zoom factor & 8 & 8 & 8 \\
sky map zoom factor & 4 & 4 & 4 \\
spindown step (Hz/s) & $\sci{5}{-12}$ &  $\sci{5}{-12}$ &  $\sci{5}{-12}$ \\
\multicolumn{4}{c}{\em Second coincidence step} \\
maximum frequency mismatch (mHz) & $5$ & $1$ & $1$ \\
minimum increase in multi-interferometer SNR (\%) & 20 & 20 & 20 \\
minimum single-interferometer SNR (\%) & 20 & 20 & 20 \\
\multicolumn{4}{c}{\em Loosely coherent followup with coherent} \\
\multicolumn{4}{c}{\em combination of data between interferometers} \\
phases sampled & 16 & 16 & 16 \\
maximum frequency mismatch (mHz) & $5$ & $1$ & $1$ \\
minimum increase in multi-interferometer SNR (\%) & 7 & 7 & 7 \\
\hline
\end{tabular}
\caption{Detection pipeline parameters.}
\label{tab:coincidence_parameters}
\end{center}
\end{table*}

\subsection{Injections and Validation}
The loosely coherent search code has undergone the same extensive review as the regular semi-coherent PowerFlux discussed earlier. In addition to strain reconstruction tests, mismatch determination and injection recovery, we verified that the passing of reconstructed injections to the next stage of the detection pipeline does not undermine detection efficiency.

The results of such analysis in a narrow band near 400~Hz can be seen in Fig.~\ref{fig:injection_recovery}. The injection recovery ratio after the first semi-coherent pass is shown with a ``+'' symbol. The circles show recovery ratio after the first loosely coherent pass, while the crosses ``$\times$'' show recovery after the second loosely coherent stage. The improvement in parameter determination is shown in Figs. \ref{fig:spindown_improvement}, \ref{fig:distance_improvement} and \ref{fig:f0_improvement}.

\begin{figure}[htbp]
\begin{center}
  \includegraphics[width=3.0in]{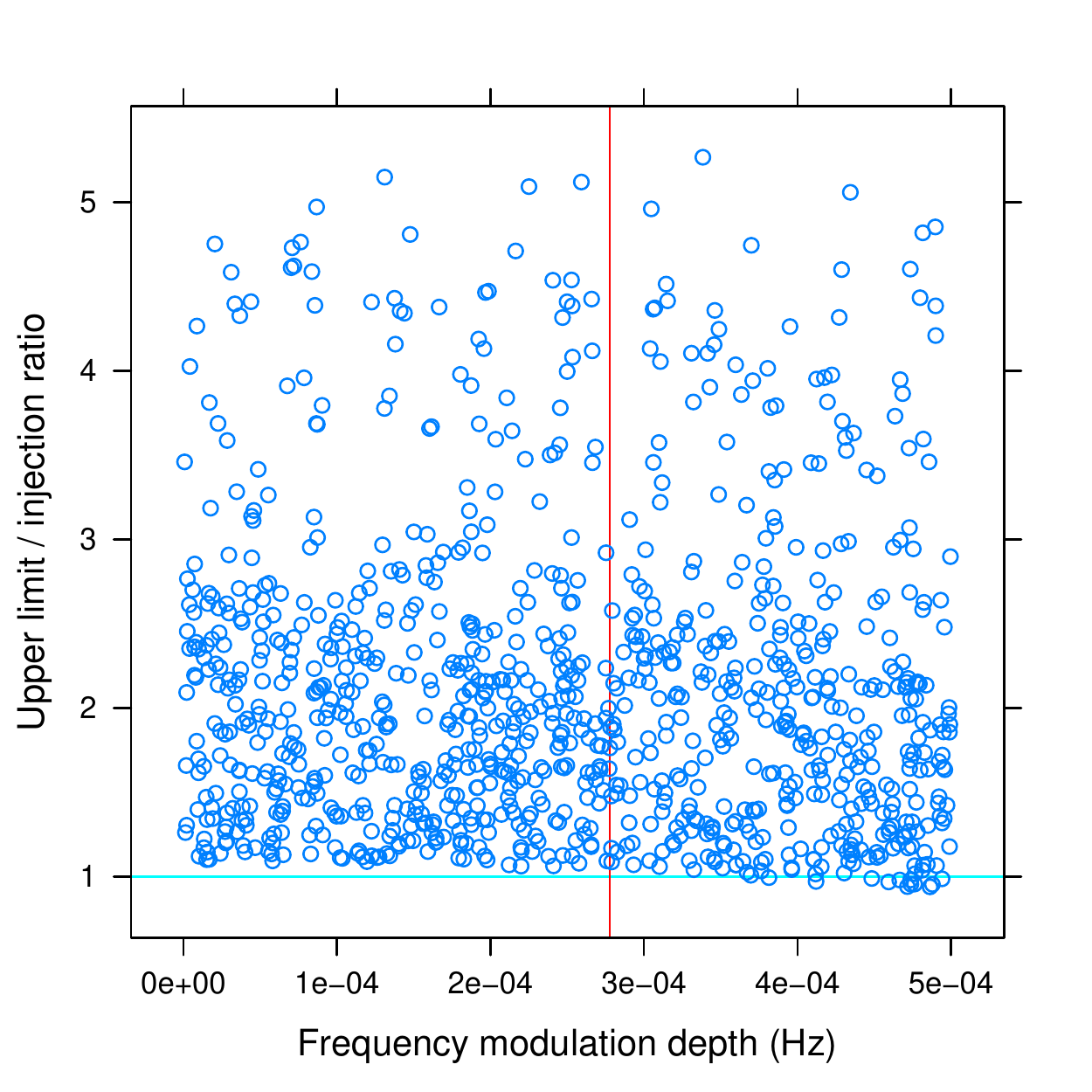}
 \caption[Upper limit versus modulation depth]{Upper limit reconstruction versus depth of periodic frequency modulation for semi-coherent search. The frequency modulation depth is shown on the x axis. Red line marks 280~$\mu$Hz boundary to the right of which we expect injections to start losing power. The y axis shows ratio between the upper limit and injected strain (color online).}
\label{fig:fmod7_single}
\end{center}
\end{figure}

\begin{figure}[htbp]
\begin{center}
  \includegraphics[width=3.0in]{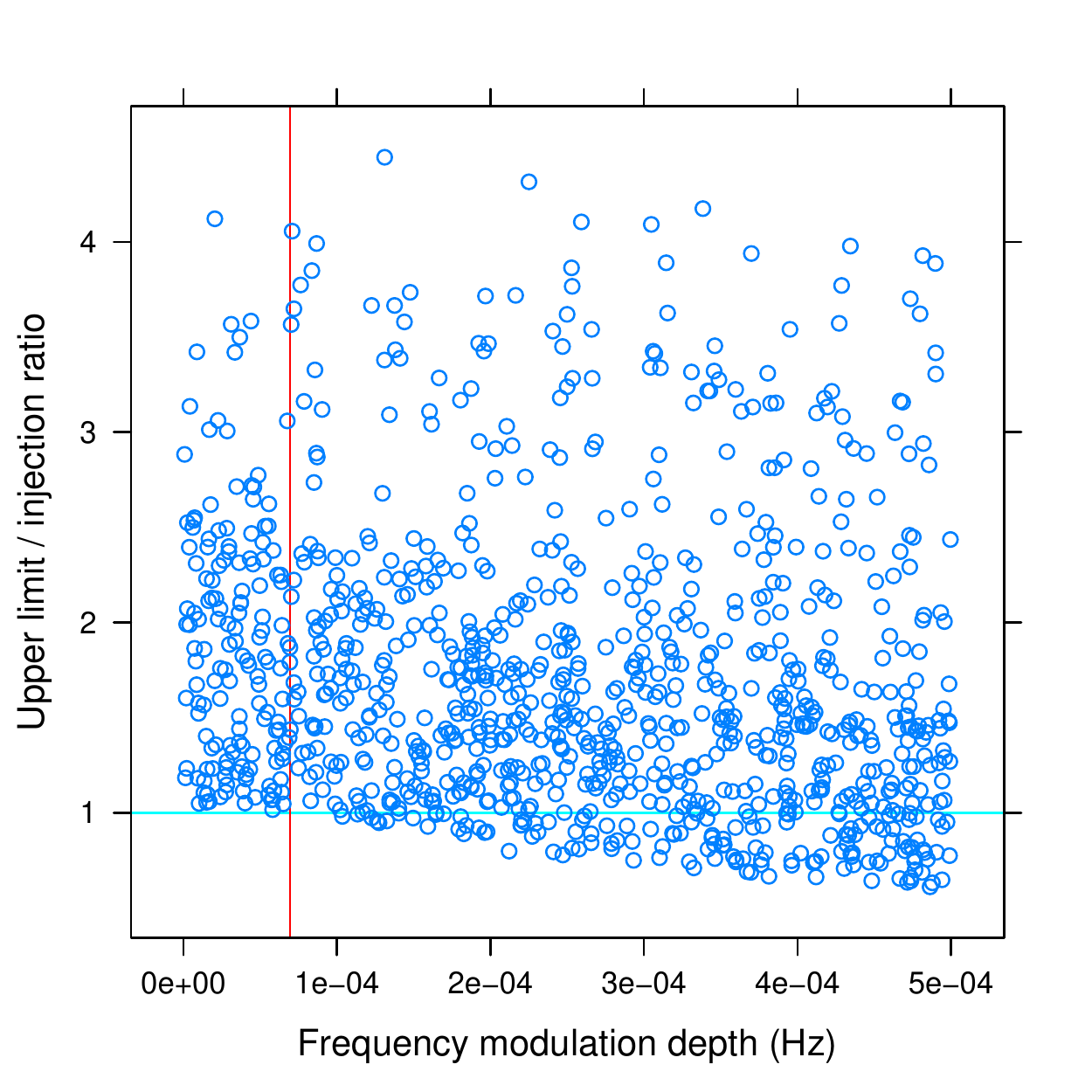}
 \caption[Upper limit versus modulation depth]{Upper limit reconstruction versus depth of periodic frequency modulation for loosely coherent search with parameter $\delta=\pi/2$. The frequency modulation depth is shown on the x axis.  Red line marks 70~$\mu$Hz boundary to the right of which we expect injections to start losing power. The y axis shows ratio between the upper limit and injected strain (color online). }
\label{fig:fmod7_pi_2}
\end{center}
\end{figure}

We have also run a simulation to determine whether the loosely coherent followup preserves the robustness to deviations from the ideal signal model that we obtain with a regular semi-coherent code. Figure \ref{fig:fmod7_single} shows the results of simulation where we applied an                                                                                            
additional sinusoidal frequency modulation to the signal.  We                                                                                             
considered frequency modulations with periods above 2 months. 
 Figure \ref{fig:fmod7_pi_2} shows results for the loosely coherent pipeline. The red line marks the amplitude of frequency modulation where we                                                                                        
had predicted we would start to see significant signal loss, based on rough estimates of how much power is expected to ``leak'' into                                                                   
adjacent frequency bins.  For the semi-coherent search the tolerance is 280~$\mu$Hz, while for the loosely coherent search it is 70~$\mu$Hz. 

\section{Results}
\label{sec:results}

\begin{figure}[htbp]
\includegraphics[width=3in]{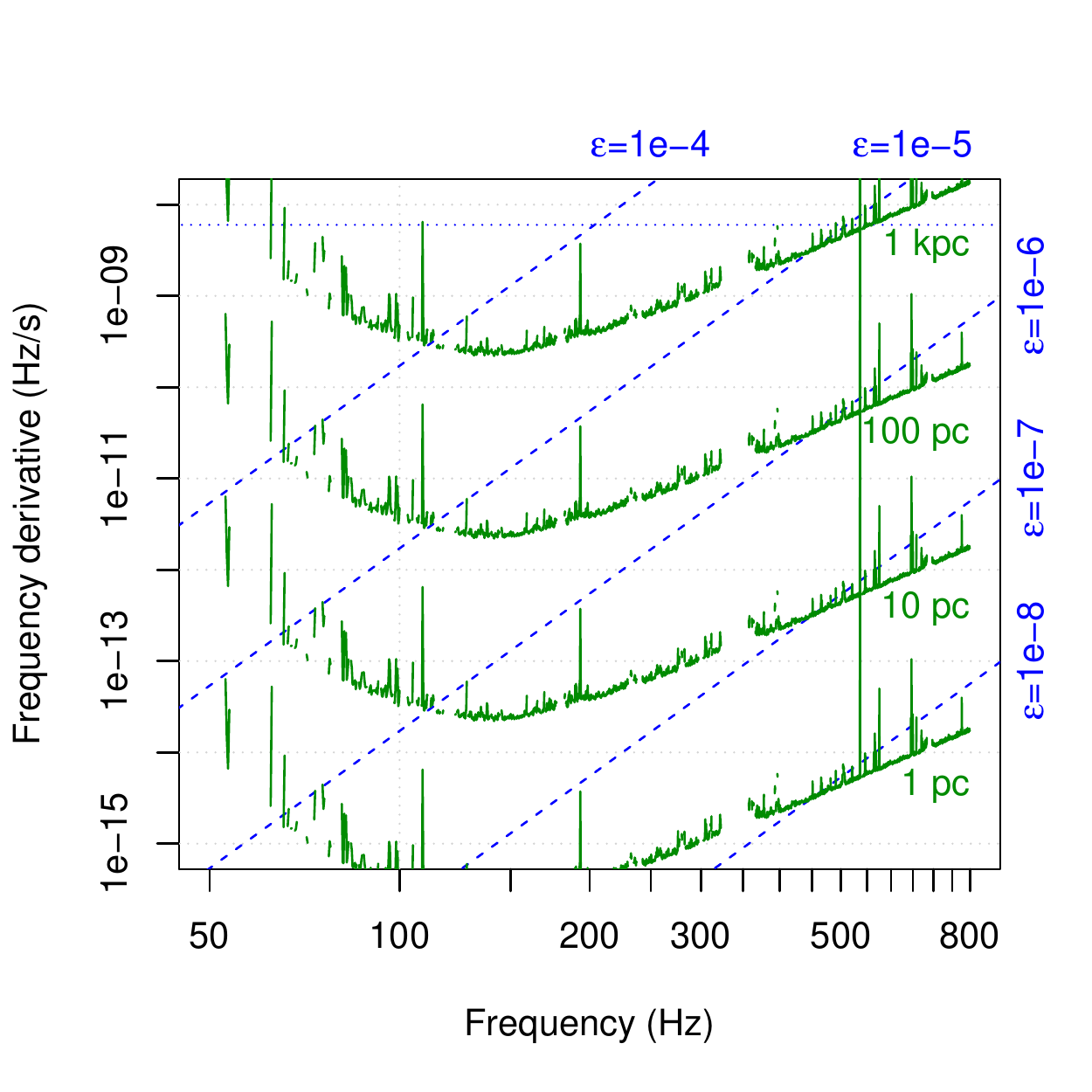}
\caption[Spindown range]{
\label{fig:spindown_range}
Range of the PowerFlux search for neutron stars
spinning down solely due to gravitational radiation.  This is a
superposition of two contour plots.  The green solid lines are contours of the maximum distance at which a neutron
star could be detected as a function of gravitational-wave frequency
$f$ and its derivative $\dot{f}$.  The dashed lines 
are contours of the corresponding ellipticity
$\epsilon(f,\dot{f})$. The fine dotted line marks the maximum spindown searched. Together these quantities tell us the
maximum range of the search in terms of various populations (see text
for details) (color online).  }
\end{figure}

PowerFlux produces 95\% confidence level upper limits for individual templates, where each template                                                                                               
represents a particular value of frequency, spindown, sky location and polarization. The results are maximized over several parameters, and a correction factor is applied to account for possible mismatches of real signal with sampled parameters. Figure \ref{fig:full_s5_upper_limits} shows the resulting upper limits maximized over the analyzed spindown range, over the sky and, for the upper set of curves, over all sampled polarizations. The lower set of curves shows the upper limit for circular polarization alone.
The uncertainty of these values is below $15$\% \cite{S5_calibration}, dominated by systematic and statistical calibration errors. The numerical data for this plot can be obtained separately \cite{data}.

The solid blue points denote values for which we found evidence of non-Gaussian behaviour in the underlying data. For these, we do not claim a specific confidence bound. The regions near harmonics of 60 Hz power line frequency are shown as circles. In addition, a small portion of the sky near each ecliptic pole has been excluded from the search, as these regions are susceptible to contamination from stationary
instrumental spectral lines. The excluded portion consists of sky templates where frequency shifts due to Doppler modulation and spindown are close to each other for a significant fraction of input data \cite{EarlyS5Paper}. This is similar to the S parameter veto described in \cite{S4IncoherentPaper}, but takes into account varying noise level in input SFTs. The fraction of excluded sky starts at about $1$~\% at 50~Hz and decreases as $f^{-2.15}$ with deviations due to wideband instrumental artifacts.

Figure \ref{fig:spindown_range} provides an easy way to judge the astrophysical range of the search. We have computed the implied spindown solely due to gravitational emission at various distances, as well as corresponding ellipticity curves. This follows formulas in paper \cite{S4IncoherentPaper}. For example, at the highest frequency sampled, assuming ellipticity of $\sci{3.3}{-6}$ (which is well under the maximum limit in \cite{crust_limit}) we can see as far as 425 parsecs.

In each search band, including regions with detector artifacts and without restrictions on sky position, the followup pipeline described                                                                                                
in section \ref{sec:followup} was applied to outliers satisfying the initial                                                                                                  
coincidence criteria. The statistics are as follows: the second stage received 9855 outliers, out of which only 619 survived to the third stage of followup, which reduced them to 47 outliers. They are summarized in Table \ref{tab:outliers} which lists only one outlier for each frequency of interest. The frequency is specified relative to GPS time 846885755, which corresponds to the middle of the S5 run.

\begin{table*}[htbp]
\begin{center}
\begin{tabular}{D{.}{.}{6}rD{.}{.}{2}D{.}{.}{2}l}\hline
\multicolumn{1}{c}{Frequency} & \multicolumn{1}{c}{Spindown} &  \multicolumn{1}{c}{RA (J2000)}  & \multicolumn{1}{c}{DEC (J2000)} & Description \\
\multicolumn{1}{c}{Hz}	&  \multicolumn{1}{c}{Hz/s} & \multicolumn{1}{c}{degrees} & \multicolumn{1}{c}{degrees} & \\
\hline \hline
63.391111 & $\sci{-9.15}{-10}$ &  53.96 &  20.60 & Electromagnetic interference in L1\\
70.883403 & $\sci{-9.10}{-10}$ &  34.99 & -13.06 & Line in L1 from controls/data acquisition system\\
70.884306 & $\sci{-7.65}{-10}$ &  22.46 &  -9.55 & Line in L1 from controls/data acquisition system\\
80.006389 & $\sci{-2.80}{-10}$ & 247.89 & -19.86 & 16 Hz harmonic from data acquisition system \\
97.772778 & $\sci{-1.85}{-09}$ & 59.39 &   1.12 & Line in L1 from controls/data acquisition system\\
97.787569 & $\sci{-2.25}{-10}$ & 355.08 &  -1.94 & Line in L1 from controls/data acquisition system\\
108.857569 & $\sci{-2.00}{-11}$ &177.90 & -32.84 & Hardware injection of simulated signal (ip3) \\
128.057083 & $\sci{-1.81}{-09}$ &229.39 &  15.47 & 16 Hz harmonic from data acquisition system \\
180.178056 & $\sci{-2.68}{-09}$ &319.38 &  29.16 & 60 Hz harmonic \\
193.323333 & $\sci{-2.08}{-09}$ &345.78 & -27.19 & Hardware injection of simulated signal (ip8) \\
566.049375 & $\sci{1.50}{-11}$ &  81.27 &  42.29 & Suspension wire resonance in H1\\
568.077708 & $\sci{-1.84}{-09}$ &283.47 & -61.01 & Suspension wire resonance in H1 \\
575.163542 & 0                 & 215.24 &   3.47 & Hardware injection of simulated signal (ip2) \\
\hline
\end{tabular}
\caption[Outliers that passed detection pipeline]{Outliers that passed detection pipeline. Only the highest-SNR outlier is shown for each hardware injection and 60 Hz harmonic.}
\label{tab:outliers}
\end{center}
\end{table*}


\begin{table*}[htbp]
\begin{center}
\begin{tabular}{lD{.}{.}{5}rD{.}{.}{2}D{.}{.}{2}}\hline
Name & \multicolumn{1}{c}{Frequency} & \multicolumn{1}{c}{Spindown} & \multicolumn{1}{c}{RA (J2000)} & \multicolumn{1}{c}{DEC (J2000)} \\
 & \multicolumn{1}{c}{Hz} & \multicolumn{1}{c}{Hz/s} & \multicolumn{1}{c}{degrees} & \multicolumn{1}{c}{degrees} \\
\hline \hline
ip2   &  575.16356  & $\sci{-1.37}{-13}$   &     215.26   &    3.44 \\
ip3   &  108.85716 & $\sci{-1.46}{-17}$   &      178.37  &   -33.44  \\
ip8   & 193.48479 & $\sci{-8.65}{-09}$   &       351.39  &   -33.42   \\
\end{tabular}
\caption[Parameters of hardware injections]{Parameters of hardware-injected simulated signals detected by PowerFlux (epoch GPS 846885755).}
\label{tab:injections}
\end{center}
\end{table*}

Most of the 47 remaining outliers are caused by three simulated pulsar signals injected into
 the instrument as test signals. Their parameters are shown in Table \ref{tab:injections}. The signal {\tt ip8} lay outside the sampled spindown range, but was loud enough to generate an outlier at an offset from the true location and frequency. The spindown values of {\tt ip2} and {\tt ip3} are very close to $0$ and were detected in the first few templates. 

Several techniques were used to identify outlier causes. During S5 there was a general effort to identify problematic areas of frequency space and instrumental sources of the contamination. Noise lines were identified by previously performed searches \cite{S5EH, EarlyS5Paper, LSC-Stochastic} as well as the search described in this paper. In addition, a dedicated analysis code ``FScan''~\cite{FScan} was created specifically for identification of instrumental artifacts. Problematic noise lines were recorded, and monitored throughout S5. Another technique used was the calculation of the coherence between the interferometers' output channel and physical environment monitoring channels. In S5 the coherence was calculated as monthly averages; the coherence output was then mined for statistically significant peaks.

In addition to data analysis techniques, investigations in the laboratory at the observatories provided further evidence as to the origin of noise lines. Portable magnetometers were used to find electrical sources of noise. Measurements of the noise coming from power supplies and cooling fans in electronics racks also helped to identify a number of noise lines.

The 47 remaining outliers were investigated and were all traced to known instrumental artifacts or hardware injections.
Hence the search has not revealed a true continuous gravitational wave signal.

\section{Conclusions}

We have performed the most sensitive all-sky search to date for continuous gravitational waves 
in the range 50-800~Hz. At the highest frequencies we are sensitive to neutron stars with an equatorial 
ellipticity as small as $\sci{3.3}{-6}$ as far away as $425$~pc for unfavorable spin orientations.
For favorable orientations (spin axis aligned with line of sight), we are sensitive to ellipticities
as small as $\sci{1.2}{-6}$ for the same distance and frequencies.
A detection pipeline based on a {\em loosely coherent} algorithm was applied to outliers from our search. 
This pipeline was demonstrated to be able to detect simulated signals at the upper limit level.  However, no true pulsar signals were found.

The analysis of the next set of data produced by the LIGO and Virgo interferometers (science runs S6, VSR2 and VSR3) is under way. 
This science run has an improved strain sensitivity by a factor of two at high frequencies, but
spans a shorter observation time than S5, and its data at lower frequencies are characterized by larger contaminations
of non-Gaussian noise than for S5.
Therefore, we do not expect to produce improved upper limits in the 100-300~Hz range without 
changes to the underlying algorithm until the Advanced LIGO and Advanced Virgo interferometers begin operation.

The improved sensitivity of the S6 run coupled with its smaller data volume will make 
it easier to investigate higher frequencies and larger spindown ranges, goals of 
the forthcoming S6 searches. We also look forward to results from the Virgo interferometer, 
in particular, in the frequency range below $\sim 40$~Hz which so far has been
inaccessible to LIGO interferometers.

\section{Acknowledgments}

The authors gratefully acknowledge the support of the United States
National Science Foundation for the construction and operation of the
LIGO Laboratory, the Science and Technology Facilities Council of the
United Kingdom, the Max-Planck-Society, and the State of
Niedersachsen/Germany for support of the construction and operation of
the GEO600 detector, and the Italian Istituto Nazionale di Fisica
Nucleare and the French Centre National de la Recherche Scientifique
for the construction and operation of the Virgo detector. The authors
also gratefully acknowledge the support of the research by these
agencies and by the Australian Research Council, 
the International Science Linkages program of the Commonwealth of Australia,
the Council of Scientific and Industrial Research of India, 
the Istituto Nazionale di Fisica Nucleare of Italy, 
the Spanish Ministerio de Educaci\'on y Ciencia, 
the Conselleria d'Economia Hisenda i Innovaci\'o of the
Govern de les Illes Balears, the Foundation for Fundamental Research
on Matter supported by the Netherlands Organisation for Scientific Research, 
the Polish Ministry of Science and Higher Education, the FOCUS
Programme of Foundation for Polish Science,
the Royal Society, the Scottish Funding Council, the
Scottish Universities Physics Alliance, The National Aeronautics and
Space Administration, the Carnegie Trust, the Leverhulme Trust, the
David and Lucile Packard Foundation, the Research Corporation, and
the Alfred P. Sloan Foundation.

This document has been assigned LIGO Laboratory document number \texttt{LIGO-P1100029-v36}.

\newpage

\begin{thebibliography}{99}

\def\etal{{\it et al.}}

\bibitem{S4IncoherentPaper} 
B.~Abbott \etal\ (LIGO Scientific Collaboration), 
Phys.\ Rev.\ D \textbf{77}, 022001 (2008).

\bibitem{EarlyS5Paper}
 B.~P.~Abbott \etal\ (LIGO Scientific Collaboration), Phys.\ Rev.\ Lett.\ \textbf{102}, 111102 (2009).


\bibitem{S2TDPaper} 
B.~Abbott \etal\ (LIGO Scientific Collaboration), M.~Kramer, and A.~G.~Lyne, Phys.\ Rev.\ Lett.\ {\bf 94}, 181103 (2005).

\bibitem{S3S4TDPaper} 
B.~Abbott \etal\ (LIGO Scientific Collaboration), M.~Kramer, and A.~G.~Lyne,
Phys.\ Rev.\ D {\bf 76}, 042001 (2007).

\bibitem{S2FstatPaper} 
B.~Abbott \etal\ (LIGO Scientific Collaboration), Phys.\ Rev.\ D \textbf{76}, 082001 (2007).

\bibitem{Crab} 
B.~Abbott \etal\ (LIGO Scientific Collaboration), Astrophys.\ J.\ Lett.\ \textbf{683}, 45 (2008).



\bibitem{pulsars3} 
B.~P.~Abbott \etal\ (LIGO Scientific Collaboration and Virgo Collaboration), Astrophys.\ J.\ \textbf{713}, 671 (2010).

\bibitem{CasA} 
J.~Abadie \etal\ (LIGO Scientific Collaboration), Astrophys.\ J.\ \textbf{722}, 1504 (2010).


\bibitem{BOINC}
The Einstein@Home project is built upon the BOINC (Berkeley Open Infrastructure for Network Computing) 
architecture described at \texttt{http://boinc.berkeley.edu/}.

\bibitem{S4EH} 
B.~Abbott \etal\ (LIGO Scientific Collaboration), Phys.\ Rev.\ D {\bf 79}, 022001 (2009).

\bibitem{S5EH} 
B.~P.~Abbott \etal\ (LIGO Scientific Collaboration), Phys.\ Rev.\ D {\bf 80}, 042003 (2009).

\bibitem{loosely_coherent} 
V.~Dergachev, Class.\ Quantum Grav.\ {\bf 27}, 205017 (2010).

\bibitem{LIGO_detector} 
B.~Abbott \etal\ (LIGO Scientific Collaboration), Rep.\ Prog.\ Phys.\ \textbf{72}, 076901 (2009).

\bibitem{S5_calibration} 
J.~Abadie \etal\ (LIGO Scientific Collaboration),  Nucl.\ Instrum.\ Meth.\ A \textbf{624}, 223 (2010).



\bibitem{PowerFluxTechNote}
V.~Dergachev,
LIGO technical document LIGO-T050186 (2005), available in 
{\tt https://dcc.ligo.org/}

\bibitem{PowerFlux2TechNote}
V.~Dergachev,
LIGO technical document LIGO-T1000272 (2010), available in 
{\tt https://dcc.ligo.org/}

\bibitem{FeldmanCousins} 
G.~J.~Feldman and R.~D.~Cousins, Phys.\ Rev.\ D {\bf 57}, 3873 (1998).

\bibitem{PowerFluxPolarizationNote}
V.~Dergachev and K.~Riles, LIGO Technical Document LIGO-T050187 (2005), available in 
{\tt https://dcc.ligo.org/}


\bibitem{gen_powerflux} 
G.~Mendell and K.~Wette, Class.\ Quantum Grav.\ \textbf{25}, 114044 (2008).

\bibitem{jks} 
P.~Jaranowski, A.~Kr\'olak, and B.~F.~Schutz,
  Phys.\ Rev.\ D {\bf 58}, 063001 (1998).
%
%

%
%
%
%


%



\bibitem{cross_correllation} 
S.~Dhurandhar, B.~Krishnan, H.~Mukhopadhyay and J.~T.~Whelan, Phys.\ Rev.\ D \textbf{77}, 082001 (2008).

\bibitem{data} See EPAPS Document No. [number will be inserted by
publisher] for numerical values of upper limits.


\bibitem{crust_limit} 
C.~J.~Horowitz and K.~Kadau, Phys.\ Rev.\ Lett.\ \textbf{102}, 191102 (2009).

\bibitem{LSC-Stochastic} 
B.~P.~Abbott \etal\ (LIGO Scientific Collaboration and Virgo Collaboration)
Nature {\bf 460}, 990 (2009).

\bibitem{FScan} 
M.~Coughlin for the LIGO Scientific Collaboration and the Virgo Collaboration, Journal of Physics: Conference Series {\bf 243}, 012010 (2010).

\end{thebibliography}

\newpage
\end{document}